\documentclass[12pt,aps,prd,preprint,groupedaddress,tightenlines,nofootinbib,floatfix]{revtex4}
\usepackage[dvipdfm]{graphicx}

\newcommand{\reseteqnum}{\setcounter{equation}{0}}
\newcommand{\nn}{\nonumber}

\newcommand{\ovl}[1]{\overline{#1}}
\newcommand{\wt}[1]{\widetilde{#1}}

\newcommand{\eqn}[1]{(\ref{#1})}
\newcommand{\p}{\partial}
\newcommand{\bpsi}{{\overline{\psi}}}

\newcommand{\vev}[1]{\left\langle #1 \right\rangle}

\newcommand{\pslash}{p\kern-1ex /}
\newcommand{\Dslash}{{\cal D}\kern-1.5ex /}

\begin{document}

\begin{flushright}
{\normalsize UTHEP-609}\\
{\normalsize UTCCS-P-59}\\
\end{flushright}

\title{Non-perturbative renormalization of quark mass 
in $N_f=2+1$ QCD with the Schr\"odinger functional scheme}

\author{S.~Aoki$^{1,2}$, K.~-I.~Ishikawa$^3$, N.~Ishizuka$^{1,2}$,
T.~Izubuchi$^4$, K.~Kanaya$^1$, Y.~Kuramashi$^{1,2}$,
K.~Murano$^{1}$, Y.~Namekawa$^2$, M.~Okawa$^3$, Y.~Taniguchi$^{1,2}$,
A.~Ukawa$^2$, N.~Ukita$^2$ and T.~Yoshi\'e$^{1,2}$\\
(PACS-CS collaboration)
}
\affiliation{
$^1$Graduate School of Pure and Applied Sciences,
University of Tsukuba,
Tsukuba, Ibaraki 305-8571, Japan\\
$^2$Center for Computational Physics,
University of Tsukuba,
Tsukuba, Ibaraki 305-8577, Japan\\
$^3$Graduate School of Science, Hiroshima University, Higashi-Hiroshima,
Hiroshima 739-8526, Japan\\
$^4$Riken BNL Research Center, Brookhaven National Laboratory, Upton,
New York 11973, USA
}

%email-address{
%saoki@het.ph.tsukuba.ac.jp
%ishikawa@theo.phys.sci.hiroshima-u.ac.jp
%ishizuka@ccs.tsukuba.ac.jp
%izubuchi@quark.phy.bnl.gov
%kadoh@ccs.tsukuba.ac.jp
%kanaya@ccs.tsukuba.ac.jp
%kuramasi@het.ph.tsukuba.ac.jp
%murano@het.ph.tsukuba.ac.jp
%namekawa@het.ph.tsukuba.ac.jp
%okawa@sci.hiroshima-u.ac.jp
%tanigchi@het.ph.tsukuba.ac.jp
%ukawa.akira.gf@un.tsukuba.ac.jp
%ukita@ccs.tsukuba.ac.jp
%yoshie@ccs.tsukuba.ac.jp
%}

\date{\today}

%=============================================
\begin{abstract}
%=============================================
 We present an evaluation of the quark mass renormalization factor for
 $N_f=2+1$ QCD. 
 The Schr\"odinger functional scheme is employed as the intermediate
 scheme to carry out non-perturbative running from the low energy region,
 where renormalization of bare mass is performed on the lattice, to deep
 in the high energy perturbative region, where the conversion to the
 renormalization group invariant mass or the  ${\ovl{\rm MS}}$ scheme is
 safely carried out.
 For numerical simulations we adopted the Iwasaki gauge action and
 non-perturbatively improved Wilson fermion action with the clover term.
 Seven renormalization scales are used to cover from low to high energy
 regions and three lattice spacings to take the continuum limit at each
 scale.
 The regularization independent step scaling function of the quark mass
 for the $N_f=2+1$ QCD is obtained in the continuum limit. 
 Renormalization factors for 
 the pseudo scalar density and the axial vector current are also
 evaluated for the same action and the bare couplings as 
 two recent large scale $N_f=2+1$ simulations;
 previous work of the CP-PACS/JLQCD collaboration, which covered
 the up-down quark mass range heavier than $m_\pi\sim 500$~MeV
 and that of PACS-CS collaboration for much lighter quark masses down 
 to $m_\pi=155$~MeV.
 The quark mass renormalization factor is used to renormalize bare
 PCAC masses in these simulations.

%=============================================
\end{abstract}
%=============================================

\maketitle

\reseteqnum
%=============================================
\section{Introduction}
%=============================================

The strong coupling constant and the quark masses constitute the 
fundamental parameters of the Standard Model.
It is an important task of lattice QCD to determine these
parameters using inputs at low energy scales such as 
hadron masses, meson decay constants and heavy quark potential
quantities.
The results can be compared with independent determinations from high
energy experiments.
%, which should provide a firm evidence of the single scale nature of QCD. 

In the course of evaluating these fundamental parameters we need the
process of renormalization in some scheme.
The ${\ovl{\rm MS}}$ scheme is one of the most popular schemes, and 
hence one would like to evaluate the running coupling constant and quark
masses through input of low energy quantities on the lattice and convert
it to the ${\ovl{\rm MS}}$ scheme.
A difficulty in this process is that the conversion is given only in a
perturbative expansion, and should be performed at much higher energy
scales than the QCD scale.
At the same time the renormalization scale $\mu$ should be kept much
smaller than the lattice cut-off to reduce lattice artifacts.
Furthermore we may need to suppress artifacts due to a finite lattice
extension $L$.  We therefore require 
\begin{eqnarray}
\frac{1}{L}\ll\Lambda_{\rm QCD}\ll\mu\ll\frac{1}{a}.
\end{eqnarray}
The practical difficulty of satisfying these inequalities in numerical 
simulations is called the window problem. 

The Schr\"odinger functional (SF) scheme
\cite{Luscher:1992an,Luscher:1993gh,Sint:1993un,Luscher:1996sc,Capitani:1998mq,Sint:2000vc,DellaMorte:2004bc,Della Morte:2005kg,Aoki:2009tf}
is designed to solve the window problem.
It has an advantage that systematic errors can be unambiguously
controlled.
A unique renormalization scale is introduced through the box size $L$ in
the chiral limit and the scheme is mass independent.
A wide range of renormalization scales can be covered by the step
scaling function (SSF) technique.
This matches our goal to obtain the coupling constant and quark masses
in the $\ovl{\rm MS}$ scheme.
The SF scheme has been applied for evaluation of the QCD coupling for
$N_f=0$ \cite{Luscher:1993gh}, $N_f=2$ \cite{DellaMorte:2004bc} and
$N_f=2+1$ \cite{Aoki:2009tf} cases.

For the quark mass renormalization factor the SF scheme has been applied
for $N_f=0$ \cite{Capitani:1998mq} and $N_f=2$
\cite{Della Morte:2005kg} QCD.
At low energy scales of $\mu\sim500$ MeV, where physical input is given,
we expect the strange quark contribution to be important in addition to
those of the up and down quarks.
Thus the aim of the present paper is to go one step further
and evaluate the quark mass renormalization factor in $N_f=2+1$ QCD.
Our goal is to renormalize the bare light quark masses and meson decay
constants evaluated in two recent large-scale $N_f=2+1$ lattice QCD
simulations employing non-perturbatively O(a) improved Wilson quark
action;
the work of CP-PACS/JLQCD Collaboration with relatively heavy pion mass
of $m_\pi\sim 500$~MeV \cite{Ishikawa:2007nn} and that of PACS-CS
collaboration with much lighter quark masses down to $m_\pi=155$~MeV
\cite{Aoki:2008sm,Kuramashi}.
%Hence the physical scale is introduced through hadrom masses given in
%these simulations.

\reseteqnum
%=============================================
\section{Our strategy}
%=============================================
\label{sec:strategy}

Our goal is to evaluate the renormalization group invariant (RGI) quark
mass $M$ given by
\begin{eqnarray}
M=\ovl{m}(\mu)\left(2b_0\ovl{g}^2(\mu)\right)^{-\frac{d_0}{2b_0}}
\exp\left(-\int_{0}^{\ovl{g}(\mu)}dg
\left(\frac{\tau(g)}{\beta(g)}-\frac{d_0}{b_0g}\right)\right),
\label{eqn:RGImass}
\end{eqnarray}
where $\ovl{g}(\mu)$ and $\ovl{m}(\mu)$ are renormalized coupling and
quark mass in some scheme at scale $\mu$.
$\beta(g)$ and $\tau(g)$ are the renormalization group (RG) function of
coupling and mass in the same scheme.
% whose perturbative expansions are given by
%\begin{eqnarray}
%&&
%\beta(g)=-g^3\left(b_0+b_1g^2+b_2g^4+\cdots\right),
%\\&&
%\tau(g)=-g^2\left(d_0+d_1g^2+d_2g^4+\cdots\right).
%\end{eqnarray}
%Coefficients are universal up to two and one loop level for $b_n$ and
%$d_n$
%\begin{eqnarray}
%&&
%b_0=\frac{1}{(4\pi)^2}\left(\frac{11}{3}N_c-\frac{2}{3}N_f\right),
%\\&&
%b_1=\frac{1}{(4\pi)^4}\left(\frac{34}{3}N_c^2
%-\left(\frac{13}{3}N_c-\frac{1}{N_c}\right)N_f\right),
%\\&&
%d_0=\frac{8}{(4\pi)^2}.
%\end{eqnarray}

Once the RGI mass $M$, which is scheme independent,  is evaluated, 
the conversion into the $\ovl{\rm MS}$ scheme can be carried out from the equation \eqn{eqn:RGImass} with the RG functions in the $\ovl{\rm MS}$ scheme.
%The renormalization scale $\mu$ is introduced through the renormalized
%coupling $g_{\ovl{\rm MS}}(\mu)$, which may be evaluated with an input
%of the RGI scale $\Lambda_{\rm QCD}$ for fixed number of flavors
%\cite{Aoki:2009tf}
%\begin{eqnarray}
%\Lambda_{\rm QCD}=
%\mu\left(b_0\ovl{g}(\mu)\right)^{-\frac{b_1}{2b_0^2}}
%\exp\left(-\frac{1}{2b_0\ovl{g}(\mu)}\right)
%\exp\left(-\int_{0}^{\ovl{g}(\mu)}dg
%\left(\frac{1}{\beta(g)}+\frac{1}{b_0g^3}-\frac{b_1}{b_0^2g}\right)\right).
%\label{eqn:lambda}
%\end{eqnarray}

We therefore first derive the renormalization factor $Z_M(g_0)$, which
converts the bare PCAC mass at a bare coupling $g_0$ to the RGI mass.
Since the step scaling function (SSF) $\sigma(u)$ for the
running coupling has already been obtained\cite{Aoki:2009tf},  an evaluation of the RGI mass
renormalization factor proceeds in the same way as given in
Ref.~\cite{Capitani:1998mq,Della Morte:2005kg}, which we  briefly
summarize here.
\begin{enumerate}
 \item We start by preparing the coupling step scaling function
       $\sigma(u)$, which gives
       the relation between the renormalized coupling constants when the
       renormalization scale is changed by a factor two,
\begin{eqnarray}
\sigma\left(u\right)=\lim_{a\to0}\left(
\left.\ovl{g}^2(2L)\right|_{u=\ovl{g}^2(L), m=0}\right).
\end{eqnarray}
       The renormalization scale is defined through a physical box size
       $L$ where the renormalized coupling $u=\ovl{g}^2(L)$ is evaluated
       in the SF scheme.

 \item We define a reference scale $L_{\rm max}$ through a renormalized
       coupling constant $u_0=\ovl{g}^2(L_{\rm max})$.
       The value of $u_0$ is arbitrary as long as it is well in the low
       energy region to suppress lattice artifacts with
       $a/L_{\rm max}\ll1$ and within the range covered by the SSF.
       We need some physical input measured separately in simulation at
       large volume to evaluate $L_{\rm max}$ in physical units.
       In this paper we shall adopt the lattice spacing determined from
       hadron masses for this purpose.

 \item  We then calculate the step scaling function (SSF) $\Sigma_P$ for
	the pseudo scalar density, which is given by the ratio of two
	renormalization factors at different renormalization scales
	$\mu=1/L$ and $1/(2L)$ at the same bare coupling $g_0$
\begin{eqnarray}
\Sigma_P\left(u,\frac{a}{L}\right)
=\left.\frac{Z_P(g_0,a/(2L) )}{Z_P(g_0,a/L)}\right|_{\ovl{g}^2(L)=u,m=0},
\label{eqn:SSF}
\end{eqnarray}
	where mass independent scheme in the massless limit is adopted.
	We used the same definition of the pseudo scalar density
	renormalization factor $Z_P$
% in the SF scheme
	as Ref.~\cite{Capitani:1998mq,Della Morte:2005kg}.

	After taking the continuum limit with three lattice spacings
\begin{eqnarray}
\lim_{a\to0}\Sigma_P\left(u,\frac{a}{L}\right)=\sigma_P(u)
\end{eqnarray}
	and performing a polynomial fit for $\sigma_P(u)$ we get the
	non-perturbative renormalization group flow of the quark mass.

 \item Since the SSF $\sigma_P(u)$ is given as a function of
       renormalized coupling $u$ we need to prepare a sequence of
       couplings $u_i$ in order to perform a non-perturbative
       running starting from $u_0=\ovl{g}^2(L_{\rm max})$.
       $u_i$'s differ by a factor two in the renormalization scale
       $L_i=2^{-i}L_{\rm max}$ and is given by solving the SSF
       $\sigma(u_{i+1})=u_i$.
       We find that we are well in the perturbative region for $i\ge5$
       \cite{Aoki:2009tf}.

 \item Taking the product of the SSF $\sigma_P(u_i)$ at each scale we
       get a factor 
\begin{eqnarray}
\frac{\ovl{m}(1/L_n)}{\ovl{m}(1/L_{\rm max})}
=w_n^{-1}={\prod_{i=1}^n\sigma_P(u_i)}
\end{eqnarray}
       which represents the running of the renormalized mass from
       $L_{\rm max}$ to $L_n$.
       We derive the renormalization factor, which converts the quark
       mass in the SF scheme at the reference scale $L_{\rm max}$ to the
       RGI mass
\begin{eqnarray}
\frac{M}{\ovl{m}(1/L_{\rm max})}
=\frac{\ovl{m}(1/L_n)}{\ovl{m}(1/L_{\rm max})}\frac{M}{\ovl{m}(1/L_n)}
=w_n^{-1}\frac{M}{\ovl{m}(1/L_n)}.
\label{eqn:Mm}
\end{eqnarray}
       The last factor ${M}/{\ovl{m}(1/L_n)}$ is evaluated
       perturbatively for $n\ge5$ with $\tau(g)$ at two-loop order and
       $\beta(g)$ at three-loop order \cite{Bode:1999sm} using
       \eqn{eqn:RGImass}.

 \item The renormalized PCAC quark mass at scale $\mu=1/L_{\rm max}$ is
       given by
\begin{eqnarray}
\ovl{m}(1/L_{\rm max})=\lim_{a\rightarrow 0}
\frac{Z_A(g_0,a/L)}{Z_P(g_0,a/L_{\rm max})}\,
  {m_{\rm PCAC}^{\rm (bare)}}(g_0).
\label{eqn:Zm}
\end{eqnarray}
       The bare PCAC quark mass $m_{\rm PCAC}^{\rm (bare)}(g_0)$ is
       obtained in some simulation at large volume carried out at a bare
       coupling $g_0$.
       We need to calculate the renormalization factor $Z_A(g_0,a/L)$
       for the axial current \cite{Luscher:1996jn,Della Morte:2005rd}
       and $Z_P(g_0,a/L_{\rm max})$ for the pseudo scalar density at the
       same bare coupling $g_0$ in this simulation.
       The box size $L$ for the axial current need not coincide with the
       box size $L_{\rm max}$ for the pseudo scalar density since the
       former is scale independent in the continuum limit.
       We finally obtain
\begin{eqnarray}
Z_M(g_0,a/L_{\rm max})=\frac{M}{\ovl{m}(1/L_{\rm max})}\ 
\frac{Z_A(g_0,a/L)}{Z_P(g_0,a/L_{\rm max})}.
\end{eqnarray}
       The factor $Z_A(g_0,a/L)$ is also used for
       renormalization of meson decay constants.

 \item As a last step, we evaluate the mass renormalization factor
  in $\ovl{\rm MS}$ scheme at a scale $\mu$: 
\begin{eqnarray}
 Z_m^{\ovl{\rm MS}}(g_0,\mu)
  =\frac{\ovl{m}_{\ovl{\rm MS}}(\mu)}{M}Z_M(g_0,a/L_{\rm max})
\end{eqnarray}
       The factor ${\ovl{m}_{\ovl{\rm MS}}(\mu)}/{M}$ is given by using
       \eqn{eqn:RGImass} with the renormalization group functions in
       $\ovl{\rm MS}$ scheme at four loops.
       The renormalization scale $\mu$ is introduced through the
       RGI scale $\Lambda_{\ovl{\rm MS}}$ evaluated in
       Ref.~\cite{Aoki:2009tf}.
       %with \eqn{eqn:lambda}.
\end{enumerate}

\reseteqnum
%=============================================
\section{Numerical setups}
%=============================================

We adopt the renormalization group improved Iwasaki gauge action and
the ${\cal O}(a)$ improved Wilson fermion action with the clover term,
whose parameters are identical to those adopted in the previous paper
for the running coupling \cite{Aoki:2009tf}:
The boundary improvement coefficients of the gauge action are set to the
tree-level values $c^P_{\rm{t}}=1$ and $c^R_{\rm{t}}=3/2$ for Iwasaki
action \cite{Takeda:2003he}.
The improvement coefficient $c_{\rm SW}$ of the clover term is known
non-perturbatively \cite{Aoki:2005et}.
The boundary improvement coefficient $\wt{c}_t$ of the fermion action is
set to the one loop value \cite{Aoki:1998qd}.
Differences reside in the Dirichlet boundary condition for the spatial
gauge link, which is set to
\begin{eqnarray}
U_k(x)|_{x_0=0}=U_k(x)|_{x_0=T}=\pmatrix{1\cr&1\cr&&1\cr}
\label{eqn:boundary}
\end{eqnarray}
and the twisted periodic boundary condition of the fermion fields in the
three spatial directions
\begin{eqnarray}
\psi(x+L\hat{k})=e^{i\theta}\psi(x),\quad
\bpsi(x+L\hat{k})=e^{-i\theta}\bpsi(x),
\end{eqnarray}
where $\theta=0.5$ \cite{Capitani:1998mq,Della Morte:2005kg}.
These conditions are known to show a good perturbative behavior
\cite{Sint:1998iq} for the quark mass renormalization factor.

For numerical simulations, the HMC algorithm is adopted for up and down
quarks and the RHMC algorithm for the strange quark.
The masses of all three quarks are set to zero.
We adopt the CPS++ code \cite{cps} and add some modification for the SF
formalism.
Simulations were carried out mainly on the T2K-tsukuba computer at
University of Tsukuba and the T2K-tokyo computer at University of Tokyo
\cite{t2k}.

\reseteqnum
%=============================================
\section{Step scaling function}
%=============================================

\subsection{Pseudo scalar density renormalization factor}
We adopt seven renormalized coupling values to cover weak
($\ovl{g}^2=1.001$) to strong ($\ovl{g}^2=3.418$) coupling regions
\cite{Aoki:2009tf},
which approximately satisfy $\ovl{g}^2_{i+1}(L) = \ovl{g}^2_i(2L)$.
For each coupling we use three box sizes $L/a=4, 6, 8$ to take the
continuum limit.
At the three lattice sizes the values of $\beta$ and $\kappa$ were tuned
to reproduce the same renormalized coupling keeping the PCAC mass to zero.
On the same parameters we evaluate the pseudo scalar density
renormalization factor ${Z_P(g_0,a/L)}$ and ${Z_P(g_0,a/(2L))}$ with
the Dirichlet boundary condition \eqn{eqn:boundary}.

We adopt the definition of the renormalization factor
 \cite{Capitani:1998mq,Della Morte:2005kg} given by
% as a ratio of two point functions
\begin{eqnarray}
Z_P(g_0,a/L)=\frac{1}{n_P}\frac{\sqrt{N_c f_1}}{f_P(L/2)},
\end{eqnarray}
where $f_P(x_0)$ and $f_1$ are two-point functions of pseudo scalar
density at bulk and at boundary given by
\begin{eqnarray}
&&
f_P(x_0)=-\frac{1}{N_f^2-1}\frac{a^3}{L^3}\sum_{\vec{x}}\vev{P^a(x){\cal O}^a},
\\&&
f_1=-\frac{1}{N_f^2-1}\frac{a^6}{L^6}\vev{{\cal O}^{\prime a}{\cal
O}^a},
\\&&
P^a(x)=\bpsi(x)\gamma_5\frac{1}{2}\tau^a\psi(x).
\label{eqn:pseudo-scalar}
\end{eqnarray}
The boundary pseudo scalar operators ${\cal O}^a$ and
${\cal O}^{\prime a}$ are written in terms of the boundary quark fields
$\zeta$, $\ovl{\zeta}$, $\zeta'$, $\ovl{\zeta}'$ \cite{Luscher:1996sc}
as
\begin{eqnarray}
{\cal O}^a=\sum_{\vec{u},\vec{v}}
\ovl{\zeta}(\vec{u})\gamma_5\frac{\tau^a}{2}\zeta(\vec{v})
,\quad
{\cal O}^{\prime a}=\sum_{\vec{u},\vec{v}}
\ovl{\zeta}'(\vec{u})\gamma_5\frac{\tau^a}{2}\zeta'(\vec{v}).
\end{eqnarray}
The normalization factor
\begin{eqnarray}
n_P=1+am+3\left(1-\cos\frac{a\theta}{L}\right)
\label{eqn:normalization}
\end{eqnarray}
is chosen to give $Z_P(g_0=0,a/L)=1$ at tree level on the lattice, where
the quark mass $am$ is written in terms of the PCAC mass as
\begin{eqnarray}
&&
am=-1-3\left(1-\cos\frac{a\theta}{L}\right)
+\sqrt{2am_{\rm PCAC}
+\sqrt{1+4a^2m_{\rm PCAC}^2+6\sin^2\frac{a\theta}{L}
+\left(3\sin^2\frac{a\theta}{L}\right)^2}}.
\nn\\
\end{eqnarray}
%which is given by solving the definition of the PCAC mass at tree level.
The normalization factor $n_P$ is calculated by setting $\theta=0.5$ and
the PCAC mass at each $(\beta,\kappa)$.

The value of the renormalization factors are listed in table
\ref{tab:beta-kappa} at each parameter together with the step scaling
function \eqn{eqn:SSF}.
In the table we notice that there is a tiny difference in the
renormalized coupling $\ovl{g}^2(L)$ between three boxes $L/a=4, 6, 8$.
We adopt the coupling at $L/a=8$ to define the renormalization scale
and the deviation is corrected perturbatively %for the SSF
\cite{Bode:1999sm}.
%\begin{eqnarray}
%&&
%\Sigma_P\left(u,\frac{a}{L}\right)
%=\Sigma_P\left(\wt{u},\frac{a}{L}\right)
%+\frac{\p\sigma_P(u)}{\p u}(u-\wt{u}),
%\\&&
%\sigma_P(u)=1+p_0u+p_1u^2+\cdots,
%\\&&
%p_0=-d_0\ln2,
%\label{eqn:p0}
%\\&&
%p_1=\frac{{d_0}^2{s_0}^2+2{b_0}{d_0}{s_0}^2+4{b_1}{d_0}{s_0}-4{b_0}{d_1}{s_0}
%-4{b_0}{d_0}{s_1}}{8{b_0}^2},
%\\&&
%s_0=2b_0\ln2,
%\\&&
%s_1=\left(2b_0\ln2\right)^2+2b_1\ln2,
%\end{eqnarray}
%where scheme dependent coefficient $d_1$ is given by
%\begin{eqnarray}
%d_1=d_0\left(0.0271+0.0105N_f\right)
%\end{eqnarray}
%for our boundary condition \eqn{eqn:boundary} and $\theta=0.5$.
Statistics of the runs are given in table \ref{tab:beta-kappa-conf}
together with values of the PCAC mass defined by
\begin{eqnarray}
&&
m_{\rm PCAC}
=\frac{\frac{1}{2}\left(\nabla_0+\nabla_0^*\right)f_A(\frac{T}{2})
+c_A\nabla_0^*\nabla_0f_P(\frac{T}{2})}{2f_P(\frac{T}{2})},
\\&&
f_A(x_0)=-\frac{1}{N_f^2-1}\frac{a^3}{L^3}\sum_{\vec{x}}
\vev{\left(A_0^{\rm imp.}\right)^a(x){\cal O}^a}
\end{eqnarray}
in terms of the improved axial vector current with non-perturbative
improvement coefficient \cite{Kaneko:2007wh}
\begin{eqnarray}
&&
A_\mu^{\rm imp.}(x)=A_\mu(x)+c_A\p_\mu P(x), \quad
c_A(g_0^2)=-0.0038g_0^2\frac{1-0.195g_0^2}{1-0.279g_0^2},
\label{eqn:axial-current}
\\&&
A_\mu^a(x)=\bpsi(x)\gamma_\mu\gamma_5\frac{1}{2}\tau^a\psi(x).
\end{eqnarray}

\subsection{Perturbative improvement and continuum limit}
We perform a perturbative improvement of the SSF before taking the
continuum limit.
For this purpose we need a perturbative evaluation of the lattice
artifact in the SSF
\begin{eqnarray}
\delta_{P}(u,a/L)=
\frac{\Sigma_{P}\left(u,{a}/{L}\right)-\sigma_{P}(u)}{\sigma_{P}(u)}.
\end{eqnarray}
For the SSF of the running coupling the artifact $\delta(u,a/L)$ was
evaluated at one loop level \cite{Takeda:2003he,Takeda2009}.
However, the magnitude of the one-loop correction was found to be large
and hence the range of applicability is limited
only for very weak coupling region for our lattice action,
revealing the importance of two-loop coefficient \cite{Aoki:2009tf}.
Since $\delta_P$ is not known at all for our setup,
instead of calculating $\delta_P$ at one and two-loop level
perturbatively we calculate SSF's directly by Monte-Carlo simulations
at very weak coupling $\beta\ge10$.
The same value of $\kappa$ is used as in Ref.~\cite{Aoki:2009tf}, which
gives almost vanishing PCAC mass there.
The results are listed in table \ref{tab:pt-beta-kappa} with the number
of configurations and the value of PCAC mass given in table
\ref{tab:pt-beta-kappa-conf}.

\begin{figure}
 \begin{center}
  \includegraphics[width=5.3cm]{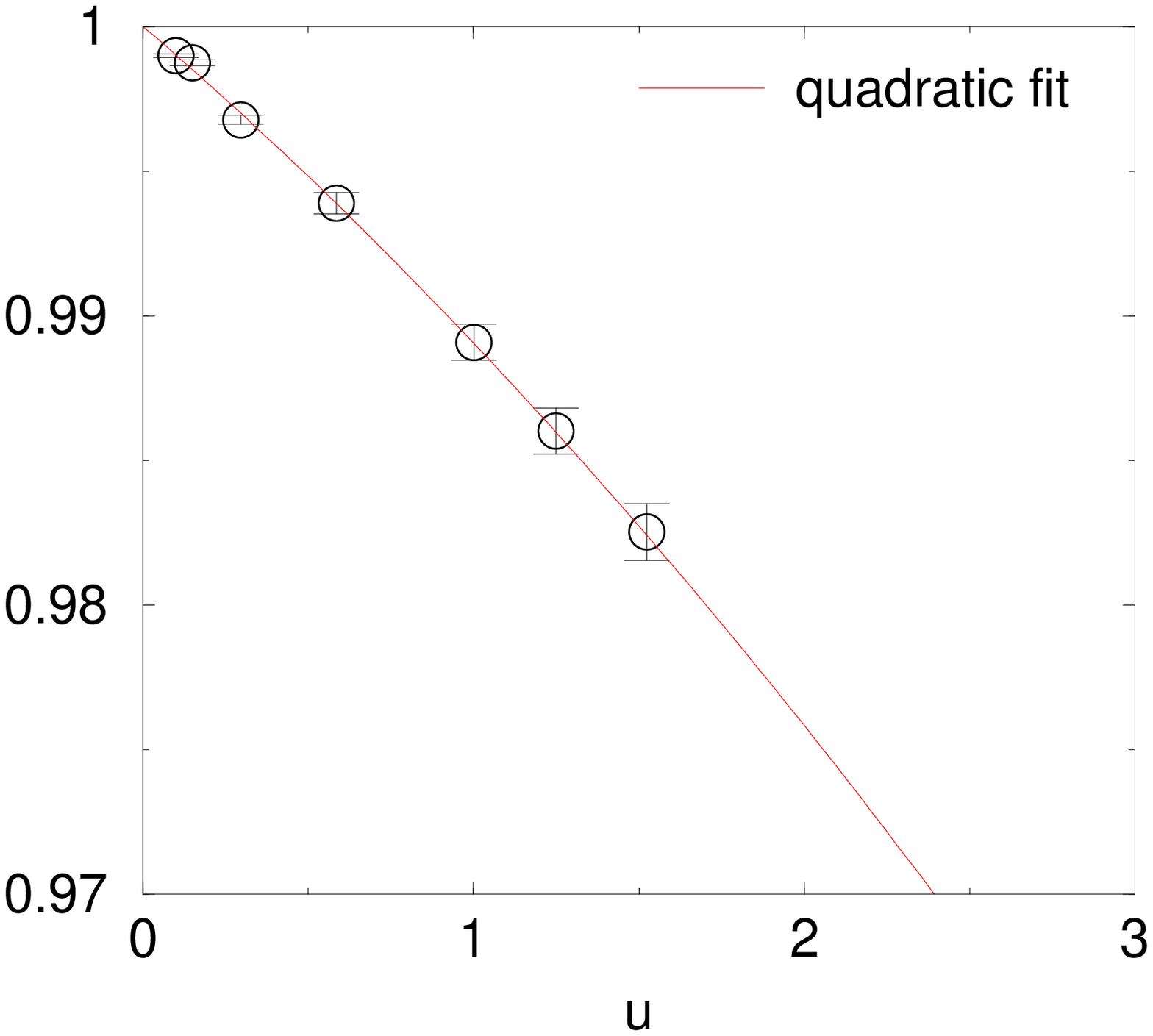}
  \includegraphics[width=5.3cm]{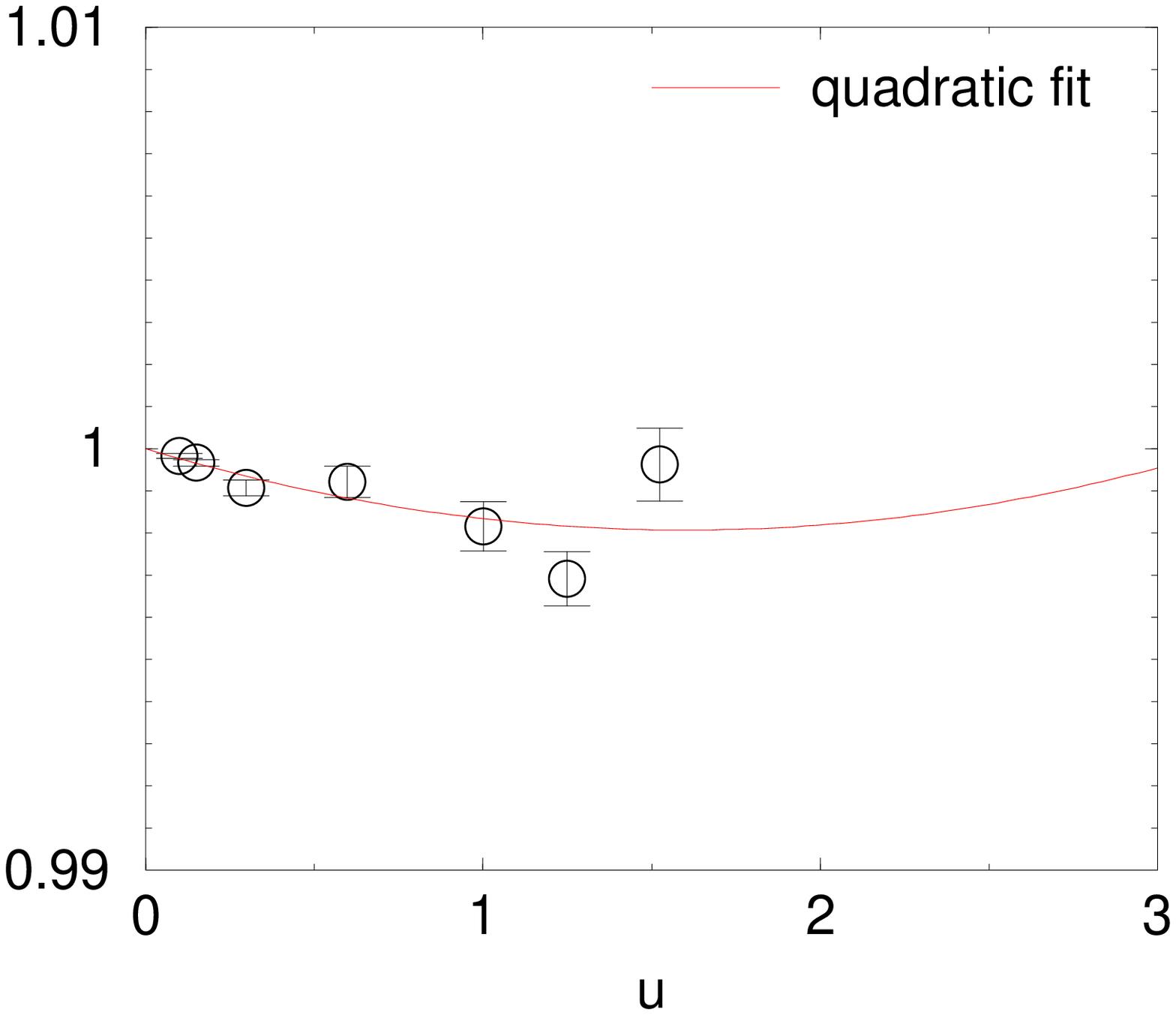}
  \includegraphics[width=5.3cm]{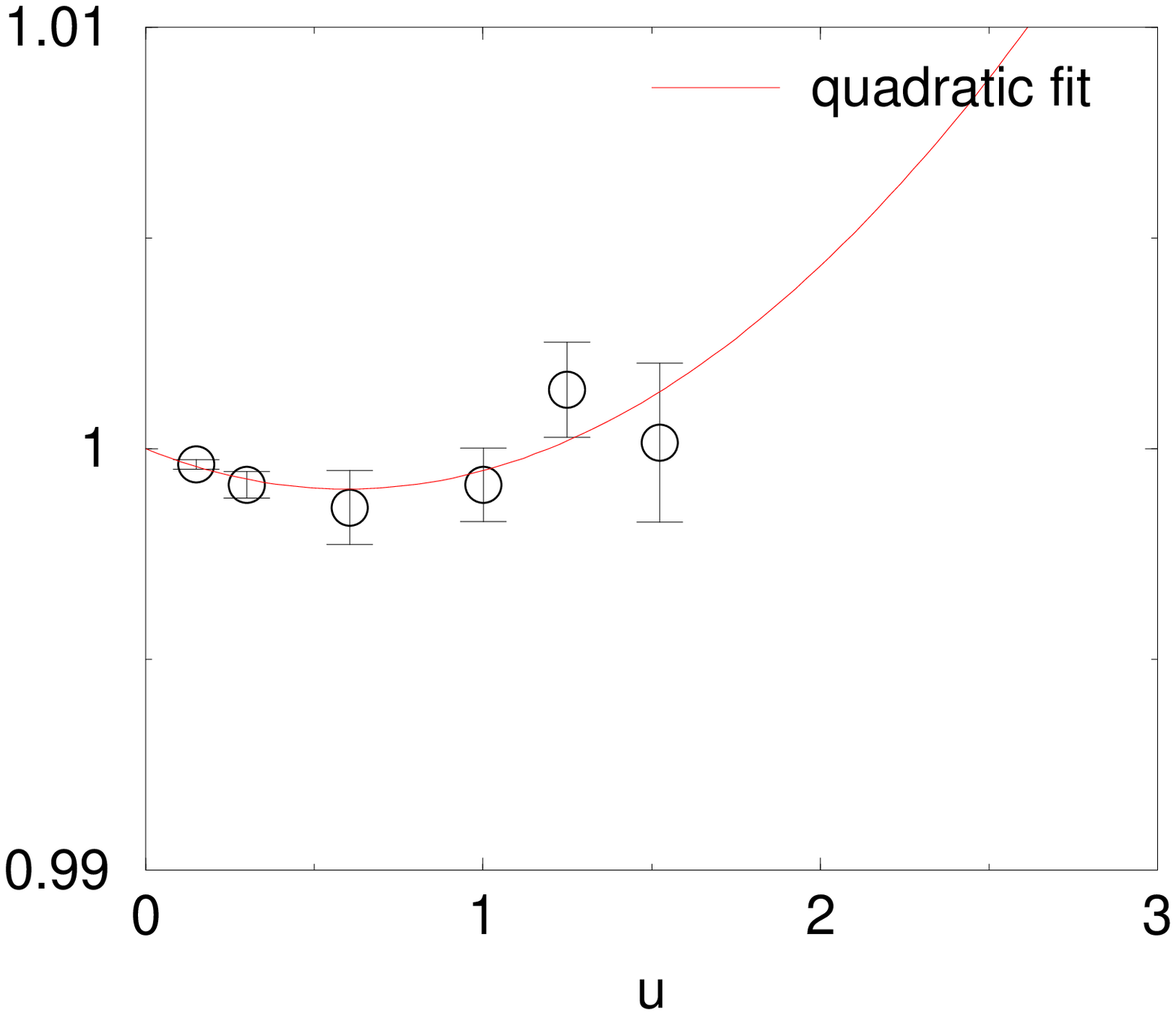}
  \caption{The polynomial fit of the discrepancy
  $\Sigma_P\left(u,{a}/{L}\right)/\sigma_P^{(2)}(u)$ at high
  $\beta\gtrsim4$.
  The fit is given for each lattice spacings $a/L=1/4$ (left), $1/6$
  (middle) and $1/8$ (right).
  Solid line is a quadratic fit.}
  \label{fig:ordera}
 \end{center}
\end{figure}

We define $\delta_{P}(u,a/L)$ by the deviation from the perturbative
SSF's $\sigma_{P}^{(2)}$ at two-loop order \cite{Bode:1999sm}
\begin{eqnarray}
\delta_{P}(u,a/L)=
\frac{\Sigma_{P}\left(u,{a}/{L}\right)-\sigma_{P}^{(2)}(u)}
{\sigma_{P}^{(2)}(u)}.
\end{eqnarray}
The deviation is fitted to a polynomial form for each $a/L$,
\begin{eqnarray}
1+\delta_{P}(u,a/L)=1+d_{1}(a/L)u+d_{2}(a/L)u^2.
\end{eqnarray}
We tried a quadratic fit using data at $u\le1.524$, which is plotted in
Fig.~\ref{fig:ordera}.
Since the quadratic fit provides a reasonable description of data
we opt to cancel the $O(a)$ contribution dividing out the SSF by the
quadratic fit
\begin{eqnarray}
\Sigma_P^{\rm (2)}\left(u,\frac{a}{L}\right)=
\frac{\Sigma_P\left(u,{a}/{L}\right)}{1+d_1(a/L)u+d_2(a/L)u^2}.
\label{eqn:pt-imp}
\end{eqnarray}
We notice that the deviation is consistent with zero within one standard
deviation for $\delta_P(u,1/8)$ even at $u\gtrsim1$.
We consider a large deviation in the quadratic fit is an artifact due to
relatively large statistical error for the data at $u\gtrsim1$ and
we therefore do not apply the present improvement for $a/L=1/8$.
Even if we apply the improvement to the SSF
$\delta_P(u,1/8)$, the quark mass renormalization factor changes only by
${\cal O}(1 \%)$, which is within statistical errors.

\begin{figure}
 \begin{center}
  \includegraphics[width=8.0cm]{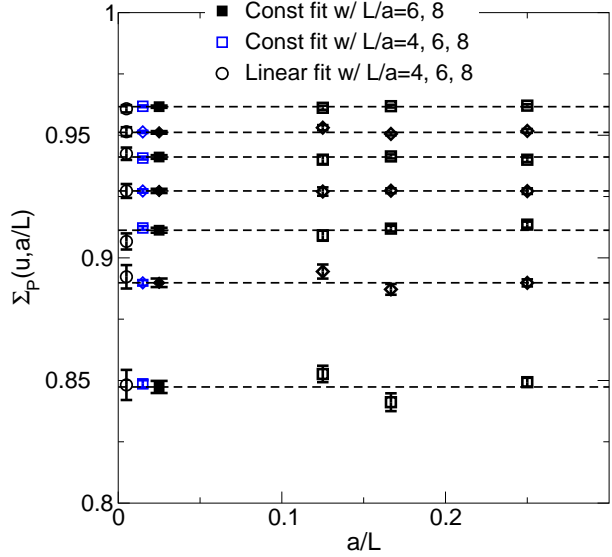}
  \caption{The SSF on the lattice with its continuum extrapolation at
  each renormalization scale.
  Three types of continuum extrapolation are tried:
  a constant extrapolation with the finest two (filled symbols) or all
  three data points (open symbols), or a linear extrapolation with all
  three data points (open circles).
  These are consistent with each other.}
  \label{fig:SSF}
 \end{center}
\end{figure}

We give the values of the perturbatively improved SSF's in table \ref{tab:ssf}
for three lattice spacings at each of the 7 renormalization scales.
% in the chiral limit.
Scaling behavior of the improved SSF is plotted in Fig.~\ref{fig:SSF}.
Almost no scaling violation is found.
We performed three types of continuum extrapolation:
a constant extrapolation with the finest two (filled symbols) or all
three data points (open symbols), or a linear extrapolation with all
three data points (open circles), which are consistent with each other.
We employed the constant fit with the finest two data points to find our
continuum value.

\begin{figure}
 \begin{center}
  \includegraphics[width=8.0cm]{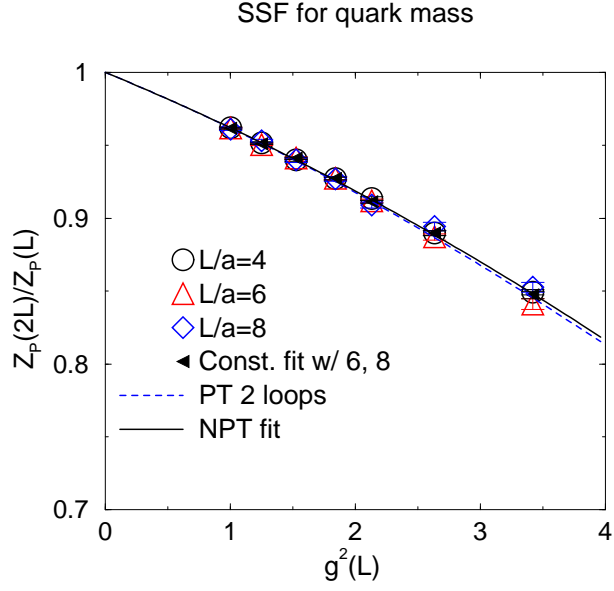}
  \caption{The RG flow of the SSF.
  Dotted line is two loops perturbative running.
  Solid line is a polynomial fit of the SSF.}
  \label{fig:SSF-RG}
 \end{center}
\end{figure}

The RG running of the continuum SSF is plotted in Fig.~\ref{fig:SSF-RG}.
A polynomial fit of the continuum SSF to third order
yields
\begin{eqnarray}
&&
\sigma_P(u)=1+p_0u+p_1u^2+p_2u^3,
\label{eqn:SSFP}
\\&&
p_1=-0.002826,\quad
p_2=0.000031
\end{eqnarray}
fixing the first coefficients to its perturbative value
$p_0=-8\ln2/(4\pi)^2$.
%\eqn{eqn:p0}
We notice that $p_1$ is consistent with its perturbative value
$p_1^{\rm (PT)}=-0.00282843$ at two loops.
The fitting function is also plotted (solid line) together with the
two loops perturbative running (dashed line).

\subsection{Non-perturbative running mass}
\begin{figure}
 \begin{center}
  \includegraphics[width=8.0cm]{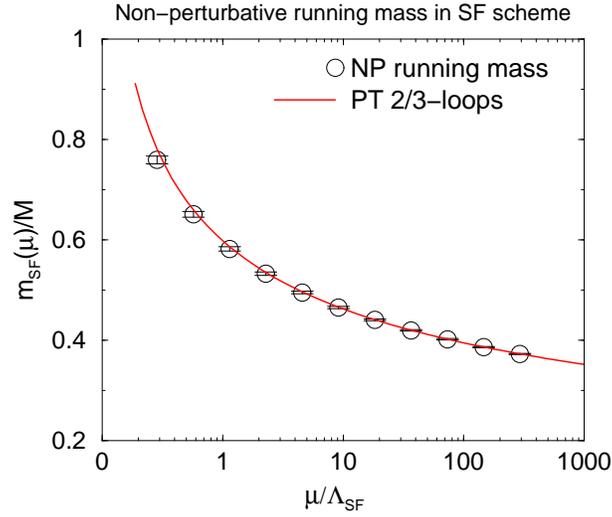}
  \caption{The non-perturbative running mass in the SF scheme.
  Solid line is a perturbative running with two and three loops RG
  function for the mass and the coupling.}
  \label{fig:running-mass}
 \end{center}
\end{figure}

Once ${\ovl{m}(1/L_{\rm max})}/M$ is evaluated from the SSF given in
\eqn{eqn:SSFP} according to the strategy in sect.\ref{sec:strategy},
we are able to derive the
non-perturbative running mass in the SF scheme according to
\eqn{eqn:Mm}.
The running mass at scale $\mu=1/L_n$ is given in units of the RGI mass
$M$ by
\begin{eqnarray}
\frac{\ovl{m}(1/L_n)}{M}
=w_n^{-1}\frac{\ovl{m}(1/L_{\rm max})}{M}.
\end{eqnarray}
The result is plotted in Fig.~\ref{fig:running-mass} as a function of
$\mu/\Lambda_{\rm SF}$.
The solid line is perturbative running given by using $\tau(g)$
at two-loop order and $\beta(g)$ at three-loop order in
\eqn{eqn:RGImass} with inputs of ${\ovl{m}(1/L_{12})}{M}$ and
$\ovl{g}(1/L_{12})$ at a very high energy scale of
$L_{12}=2^{-12}L_{\rm max}$.
We omit the horizontal error bar, which may be introduced from that of
$\Lambda_{\rm SF}$.

\reseteqnum
%=============================================
\section{$Z_P$ at low energy scale}
%=============================================

Some years ago
CP-PACS and JLQCD Collaborations jointly performed an $N_f=2+1$
simulation with the $O(a)$ improved Wilson action and the Iwasaki gauge
action \cite{Ishikawa:2007nn}.
Three values of $\beta$, $1.83$, $1.90$ and $2.05$ were adopted to take
the continuum limit and the up-down quark mass covered a rather heavy
region corresponding to $m_\pi/m_\rho=0.63-0.78$.
This project has been taken over by the PACS-CS Collaboration aiming at
simulations at the physical light quark masses
\cite{Aoki:2008sm,Kuramashi}.
At present result at a single lattice spacing $\beta=1.90$ is available 
with very light quark masses down to $m_\pi/m_\rho\approx 0.2$.

One of the purpose of this paper is to provide the non-perturbative
renormalization factor to renormalize the bare PCAC quark masses
measured in these large scale simulations.
We start by introducing a reference scale $L_{\rm max}$.
For this purpose, the renormalized coupling is evaluated at the same bare
coupling $\beta$ as in the CP-PACS/JLQCD simulations and in the chiral
limit \cite{Aoki:2009tf}.
The reference scale $L_{\rm max}$ is given by the box size we adopt in
this evaluation.
The renormalized coupling $\ovl{g}^2(L_{\rm max})$ should
not exceed our maximal value $5.13$ for the coupling SSF significantly.
We use the box size of $L/a=4$ for $\beta=1.83$ and $1.90$ to define
$L_{\rm max}$ and $L/a=6$ for $\beta=2.05$.
A physical scale is introduced into the present work so that the 
reference scale is translated into MeV units.
We employ three types of hadron masses 
$m_\pi$, $m_\rho$, $m_K$ ($m_\phi$) \cite{Ishikawa:2007nn} or
$m_\pi$, $m_K$, $m_\Omega$ \cite{Aoki:2008sm,Kuramashi}
as inputs and use the lattice spacing $a$ as an intermediate scale,
which is given in Table \ref{tab:lattice-spacing}.

Then we evaluate the pseudo scalar density renormalization factor at the
same $\beta=6/g_0^2$ and $\kappa$ as that in Ref.~\cite{Aoki:2009tf}
%the low energy scale $L_{\rm max}$
to give the denominator of the PCAC mass renormalization factor
\eqn{eqn:Zm}.
%This is done by adopting the same parameter as that in evaluation of the
%running coupling but different Dirichlet boundary condition
%\eqn{eqn:boundary} and twisted angle $\theta=0.5$.
%The same $O(a)$ improved Wilson fermion and the Iwasaki gauge action is
%adopted here.
The values of the renormalized coupling constant \cite{Aoki:2009tf} and
the renormalization factor $Z_P(g_0,a/L_{\rm max})$ are listed in
Table \ref{tab:lmax}.

\reseteqnum
%=============================================
\section{Axial current renormalization factor}
%=============================================

The remaining ingredient of the renormalization factor \eqn{eqn:Zm} at
low energy is that
%a renormalization factor 
of the axial vector current.
%\eqn{eqn:axial-current}.
We calculate the renormalization factor according to the procedure in
Ref.~\cite{Luscher:1996jn,Della Morte:2005rd} through the axial
Ward-Takahashi identity.
We adopt the renormalization condition \cite{Della Morte:2005rd}
\begin{eqnarray}
&&
Z_A^2(g_0,m)
=\wt{Z}_A\left(g_0,m,\frac{3}{2}T\right),
\\&&
\wt{Z}_A\left(g_0,m,x_0\right)
=\frac{1}{n_A}\frac{f_1}{f_{AA}^I\left(x_0,\frac{1}{3}T\right)
-2m_{\rm PCAC}\wt{f}_{PA}^I\left(x_0,\frac{1}{3}T\right)},
\label{eqn:za}
\end{eqnarray}
which is applicable to small non-vanishing PCAC mass.
In this section we suppress the $a/L_{\rm max}$ dependence while 
we explicitly write  the $m$ dependence instead.
Here $f_{AA}^I$ and $\wt{f}_{PA}^I$ are four-point functions of boundary
operators and the improved axial current \eqn{eqn:axial-current} or
the pseudo scalar density operator \eqn{eqn:pseudo-scalar} defined by
\begin{eqnarray}
&&
f_{AA}^I(x_0,y_0)=-\frac{1}{6L^6}\sum_{\vec{x}\vec{y}}
\epsilon^{abc}\epsilon^{cde}\vev{{\cal O}^{\prime d}
\left(A_0^{\rm imp.}\right)^a(x)\left(A_0^{\rm imp.}\right)^b(y){\cal O}^e},
\\&&
\wt{f}_{PA}^I(x_0,y_0)=-\frac{1}{6L^6}\sum_{\vec{x}\vec{y}}
\epsilon^{abc}\epsilon^{cde}\vev{{\cal O}^{\prime d}
P^a(x)\left(A_0^{\rm imp.}\right)^b(y){\cal O}^e}.
\end{eqnarray}
The PCAC mass here is given by an average over three time slices
\begin{eqnarray}
m_{\rm PCAC}=\frac{1}{3}\sum_{x_0=\frac{T}{2}-1}^{\frac{T}{2}+1}
\frac{\frac{1}{2}\left(\nabla_0+\nabla_0^*\right)f_A(x_0)
+c_A\nabla_0^*\nabla_0f_P(x_0)}{2f_P(x_0)}.
\end{eqnarray}
The normalization factor $n_A$ is chosen to give
$\wt{Z}_A(g_0=0,m,x_0)=1$ at tree level on the lattice.
% canceling ${\cal O}(a\theta/L)$ and ${\cal O}(am)$ effect.

As a byproduct we are able to derive the renormalization factor for the
vector current
\begin{eqnarray}
V_\mu(x)=\bpsi(x)\gamma_\mu\frac{1}{2}\tau^a\psi(x)
\end{eqnarray}
with negligible computational cost.
Here again we adopt the same definition for the renormalization factor
as in \cite{Luscher:1996jn,Della Morte:2005rd} given by
\begin{eqnarray}
&&
Z_V(g_0)=\wt{Z}_V\left(g_0,\frac{T}{2}\right),
\\&&
\wt{Z}_V\left(g_0,x_0\right)=\frac{1}{n_V}\frac{f_1}{f_V(x_0)},
\\&&
f_V(x_0)=\frac{1}{6L^6}\sum_{\vec{x}}i\epsilon^{dce}
\vev{{\cal O}^{\prime d}V^c_\mu(x){\cal O}^e}.
\end{eqnarray}
The normalization factor is given by $n_V=n_P$.
% given in \eqn{eqn:normalization} is set to have $Z_V(g_0=0)=1$ at tree
% level on the lattice.

%We adopt the same bare coupling $\beta=1.83$, $190$, $2.05$ as those in
%the large scale simulation.
There are several alternative choices in the calculation of $Z_A$:
box size $L/a$, twisting angle $\theta$ and two definitions of $Z_A$
in Ref.~\cite{Luscher:1996jn,Della Morte:2005rd} with or without
the disconnected diagrams.
Our requirement is that the ${\cal O}(a^2)$ lattice artifact is
suppressed as much as possible within a reasonable computational cost.
As a measure of the lattice artifact, we use the time dependence of
$\wt{Z}_{A/V}$, which should disappear in the continuum limit.
% by the WT identity.

We have varied the spatial box size from $L/a=6$ to $12$ until we find a
plateau in $\wt{Z}_{A/V}$ as a function of $x_0$.
Our box sizes are $6^3\times18$, $8^3\times18$, $10^3\times24$ and
$12^3\times30$.
The temporal lattice size was taken to be a multiple of six so that it is
even and the two axial currents can be placed at an equal distance
$T/(3a)$ from the boundary and from each other.
We set $T/a$ to the nearest value to $9/4\times L/a$
\cite{Della Morte:2005rd}.
The spatial twist angle is set to $\theta=0$ and $0.5$ to examine its
dependence.
The bare quark mass is tuned so that $|am_{\rm PCAC}|<0.01$
since the dependence is known to be negligible \cite{Della Morte:2005rd}
in the definition \eqn{eqn:za} for such a small mass.
The Dirichlet boundary condition \eqn{eqn:boundary} is the same as that
for $Z_P$.

\begin{figure}
 \begin{center}
  \includegraphics[width=5.cm]{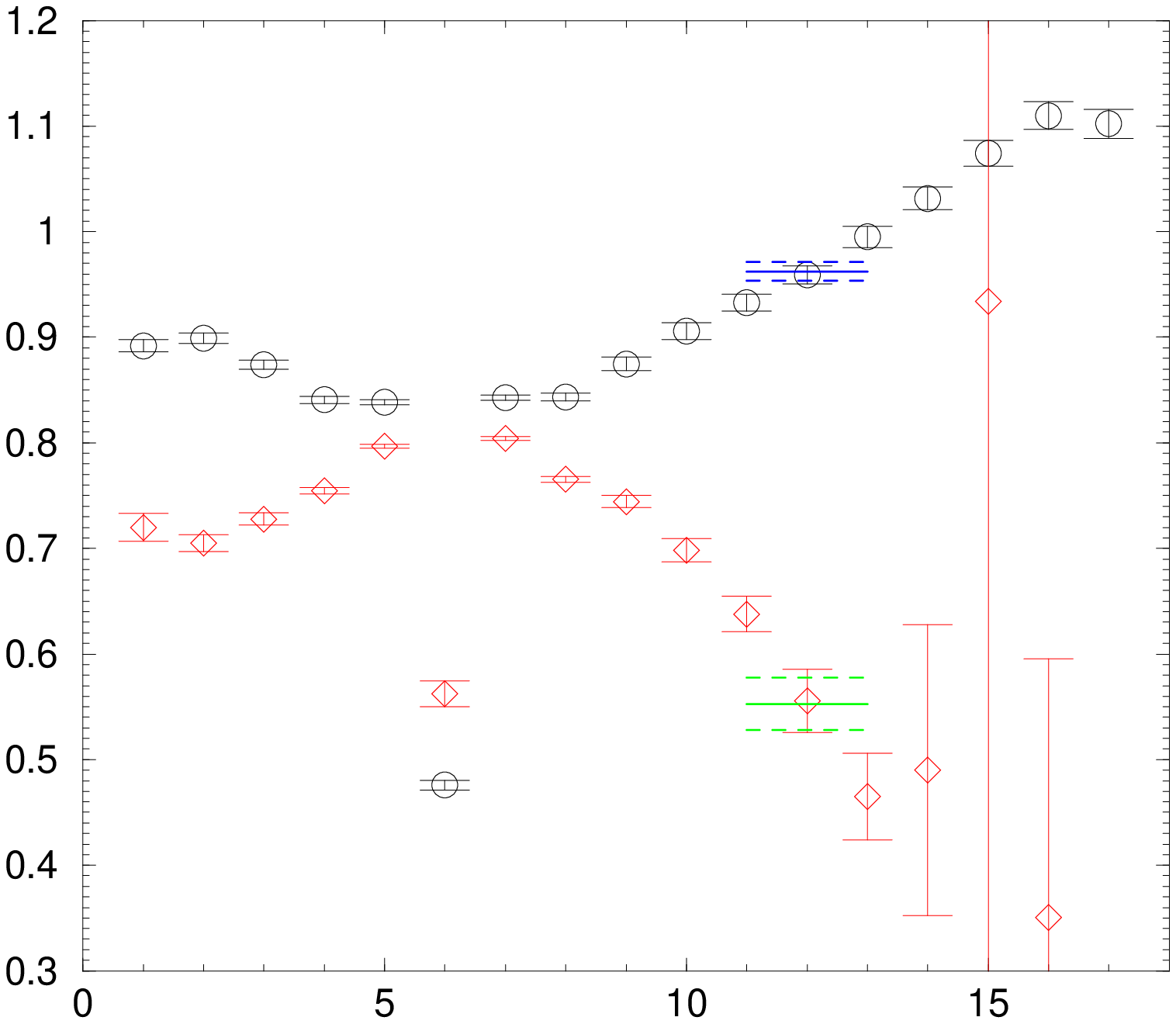}
  \includegraphics[width=5.cm]{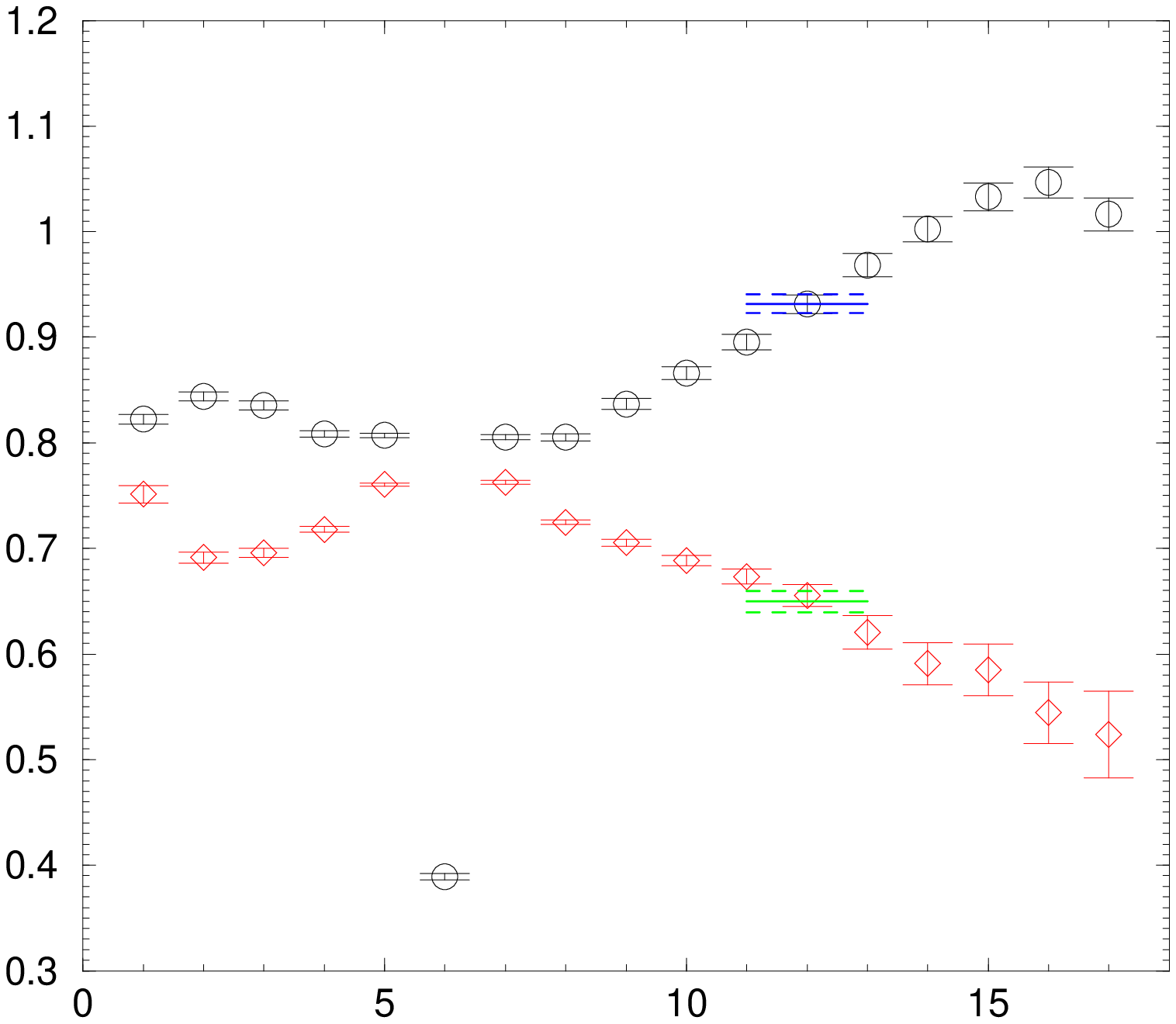}
  \includegraphics[width=5.cm]{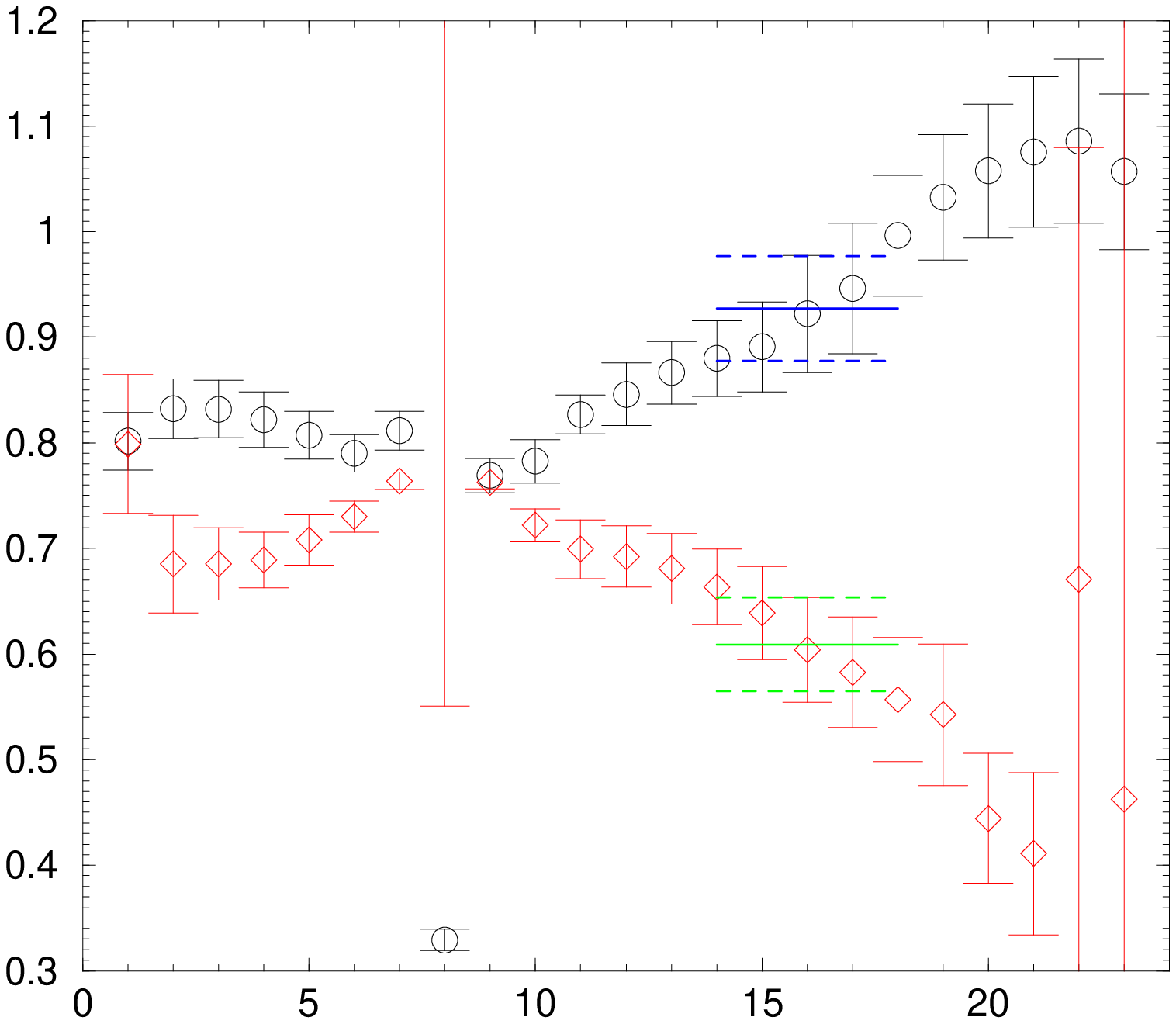}
  \includegraphics[width=5.cm]{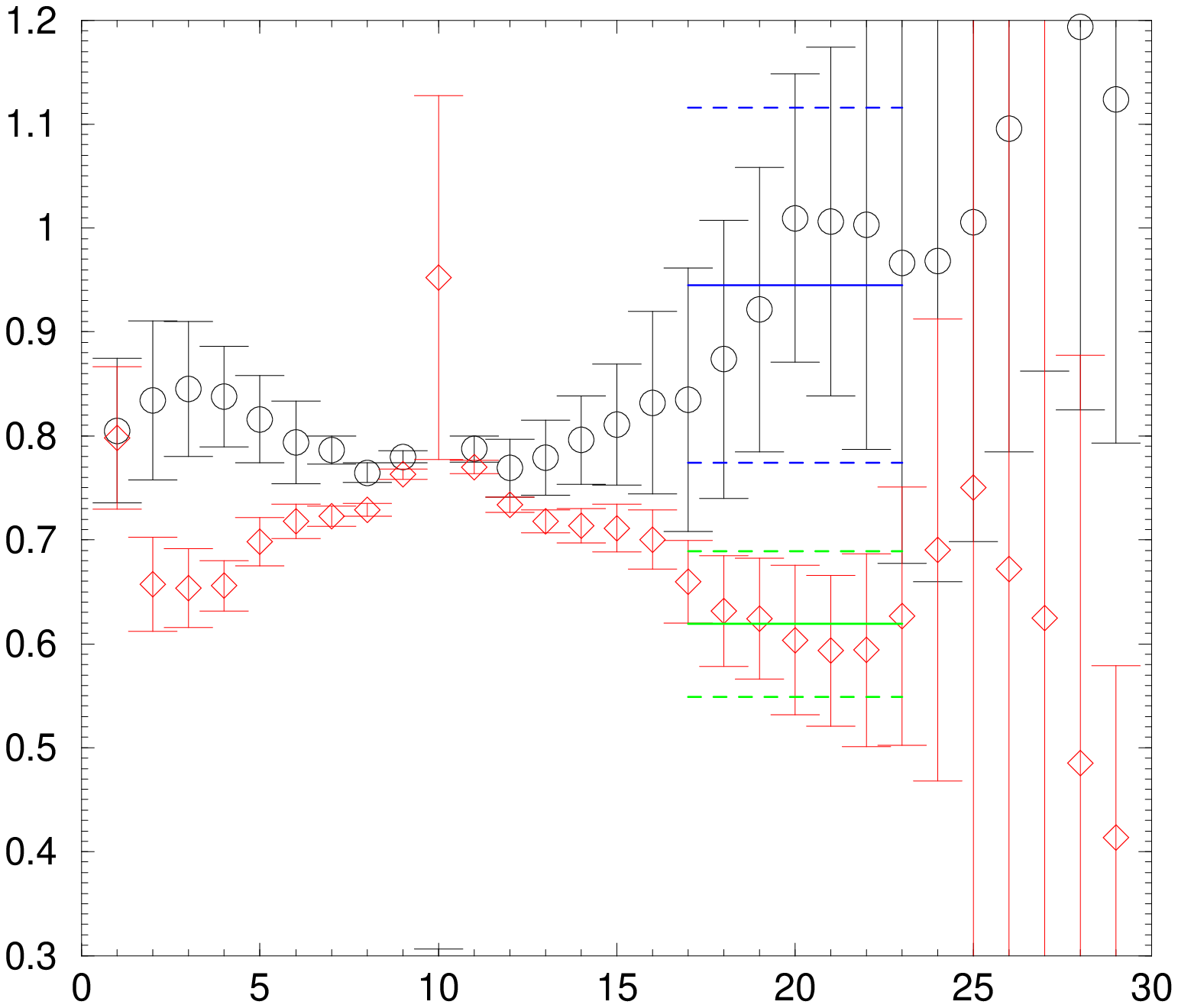}
  \caption{$x_0$ dependence of $\wt{Z}_A(g_0,m,x_0)$ at $\beta=1.83$ and
  $\theta=0.5$ for box sizes $6^3\times18$ (upper left),
  $8^3\times18$ (upper middle), $10^3\times24$ (upper right) and
  $12^3\times30$ (lower).
  Red diamonds and black circles are definition with or without the
  disconnected diagrams.
  Solid and dashed lines represent an expected plateau, which is given
  by fitting data around $x_0=2T/3$ by a constant.
  Error is estimated with the Jackknife method.
}
 \label{fig:za-b183-th05}
 \end{center}
\end{figure}

\begin{figure}
 \begin{center}
  \includegraphics[width=5.cm]{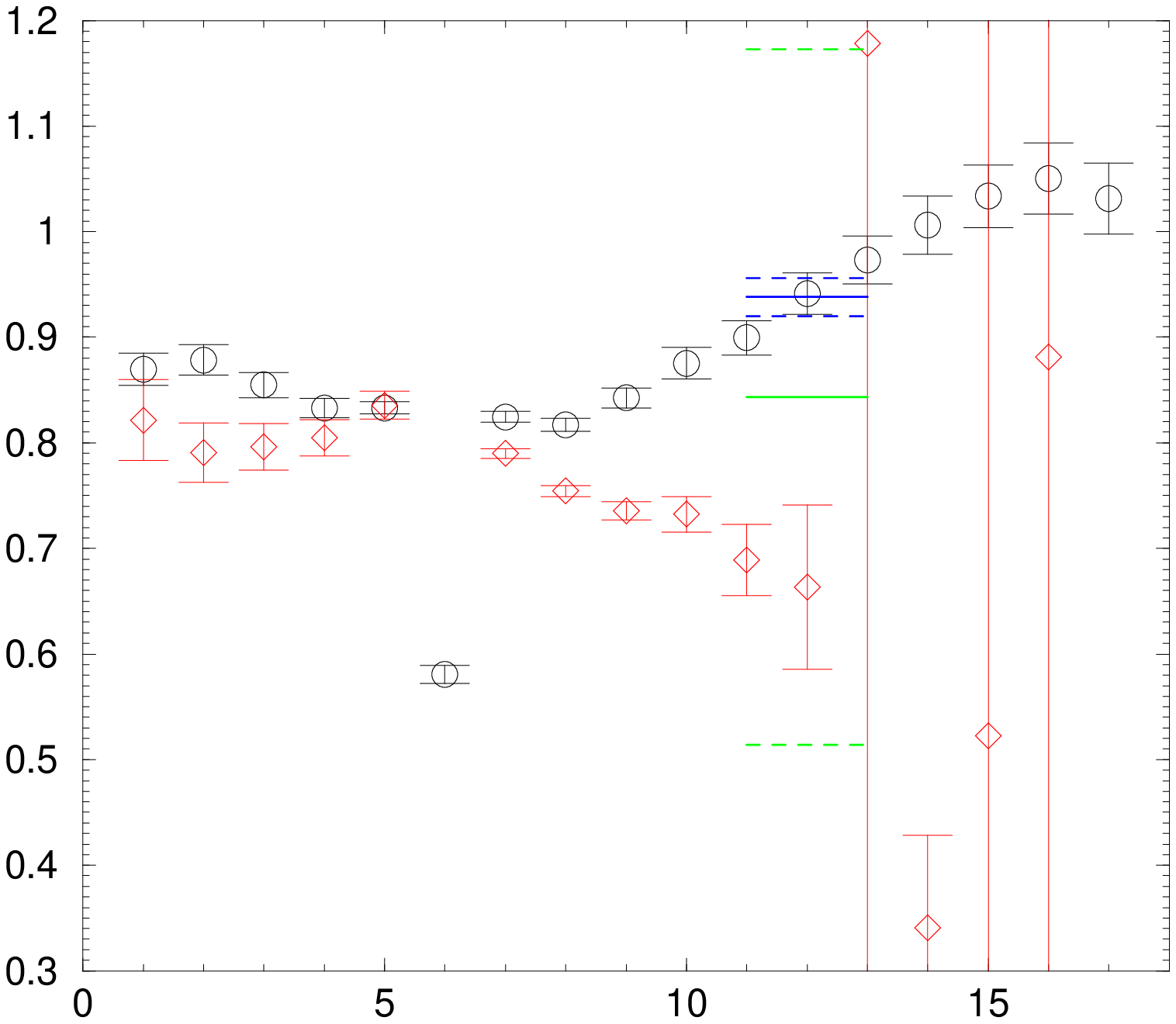}
  \includegraphics[width=5.cm]{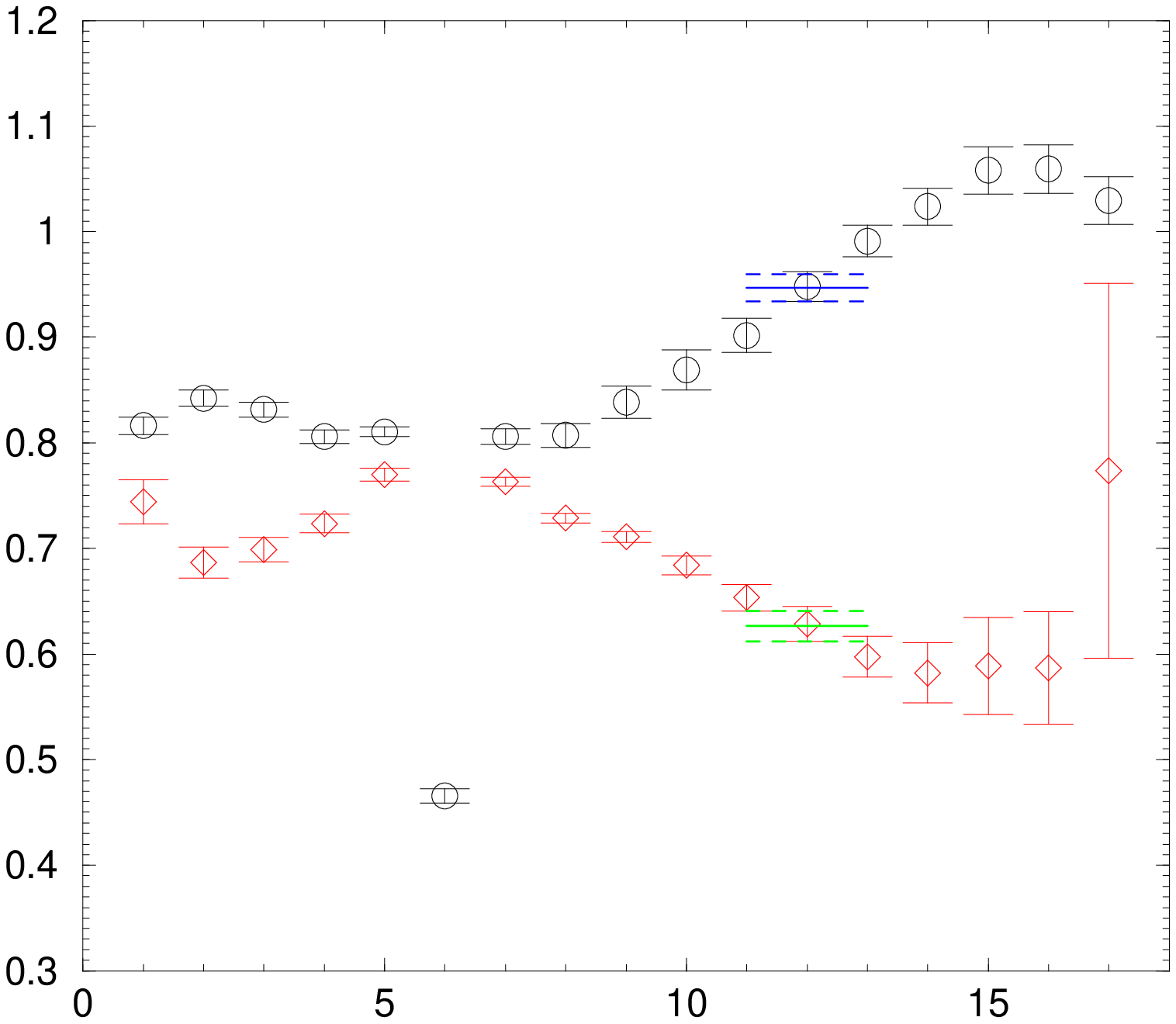}
  \includegraphics[width=5.cm]{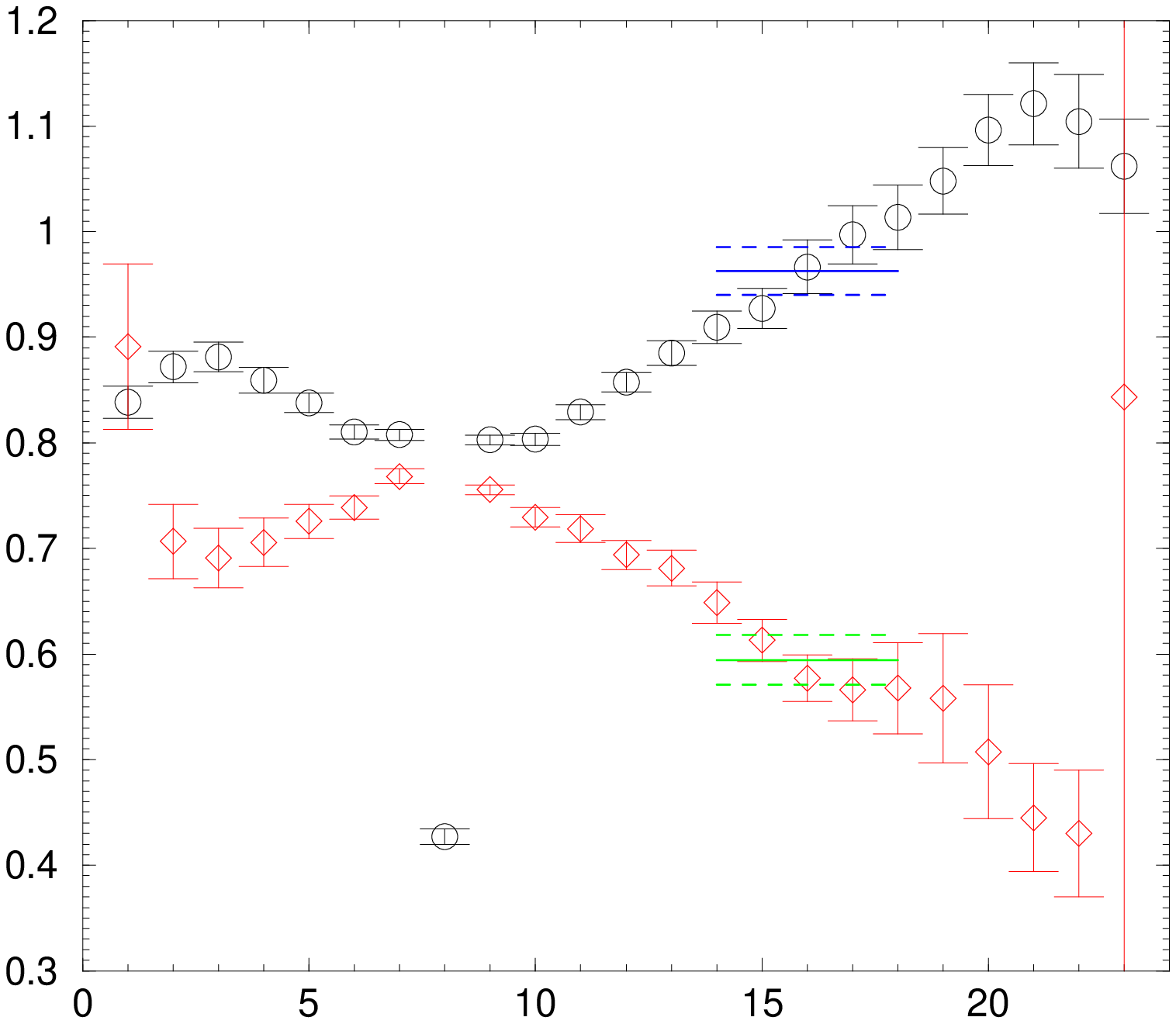}
  \caption{$x_0$ dependence of $\wt{Z}_A(g_0,m,x_0)$ at $\beta=1.83$ and
  $\theta=0$ for box sizes $6^3\times18$ (left),
  $8^3\times18$ (middle) and $10^3\times24$ (right).}
 \label{fig:za-b183-th00}
 \end{center}
\end{figure}

The time dependence of $\wt{Z}_A(g_0,m,x_0)$ is plotted for $\beta=1.83$
in Figs.~\ref{fig:za-b183-th05} ($\theta=0.5$) and \ref{fig:za-b183-th00}
($\theta=0$).
We found no plateau for both definitions with or without the
disconnected diagram for almost all box sizes.
The variation of $\theta$ does not change the conclusion.
An exception is that with the disconnected diagram on a $12^3\times30$
lattice indicated by red diamonds.
However we cannot be very confident of a plateau for two reasons.
One is that the statistical error is too large and the other is a large
discrepancy between the two definitions with or without the disconnected
diagrams, which should be a pure lattice artifact at the massless
point.

\begin{figure}
 \begin{center}
  \includegraphics[width=5.cm]{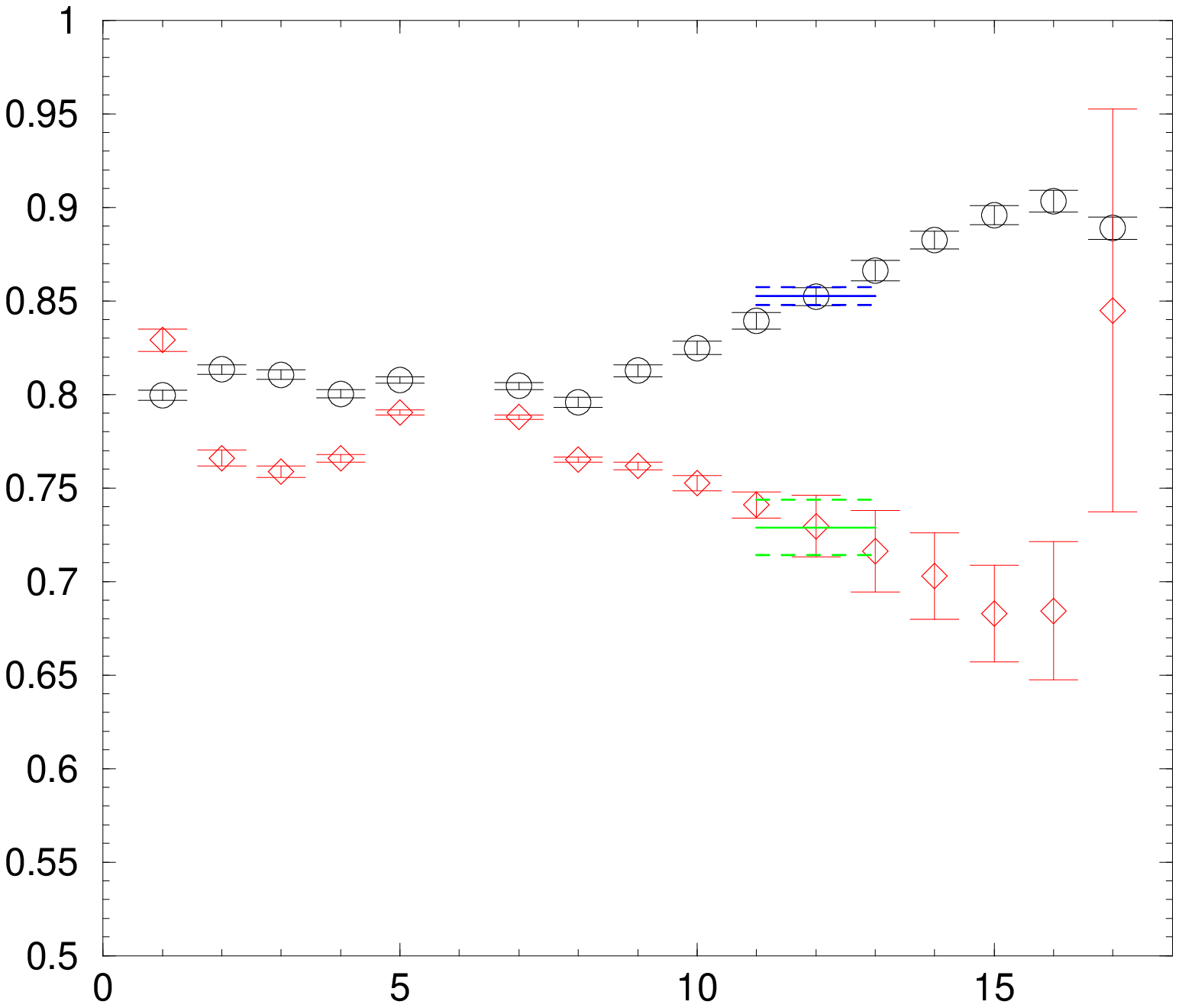}
  \includegraphics[width=5.cm]{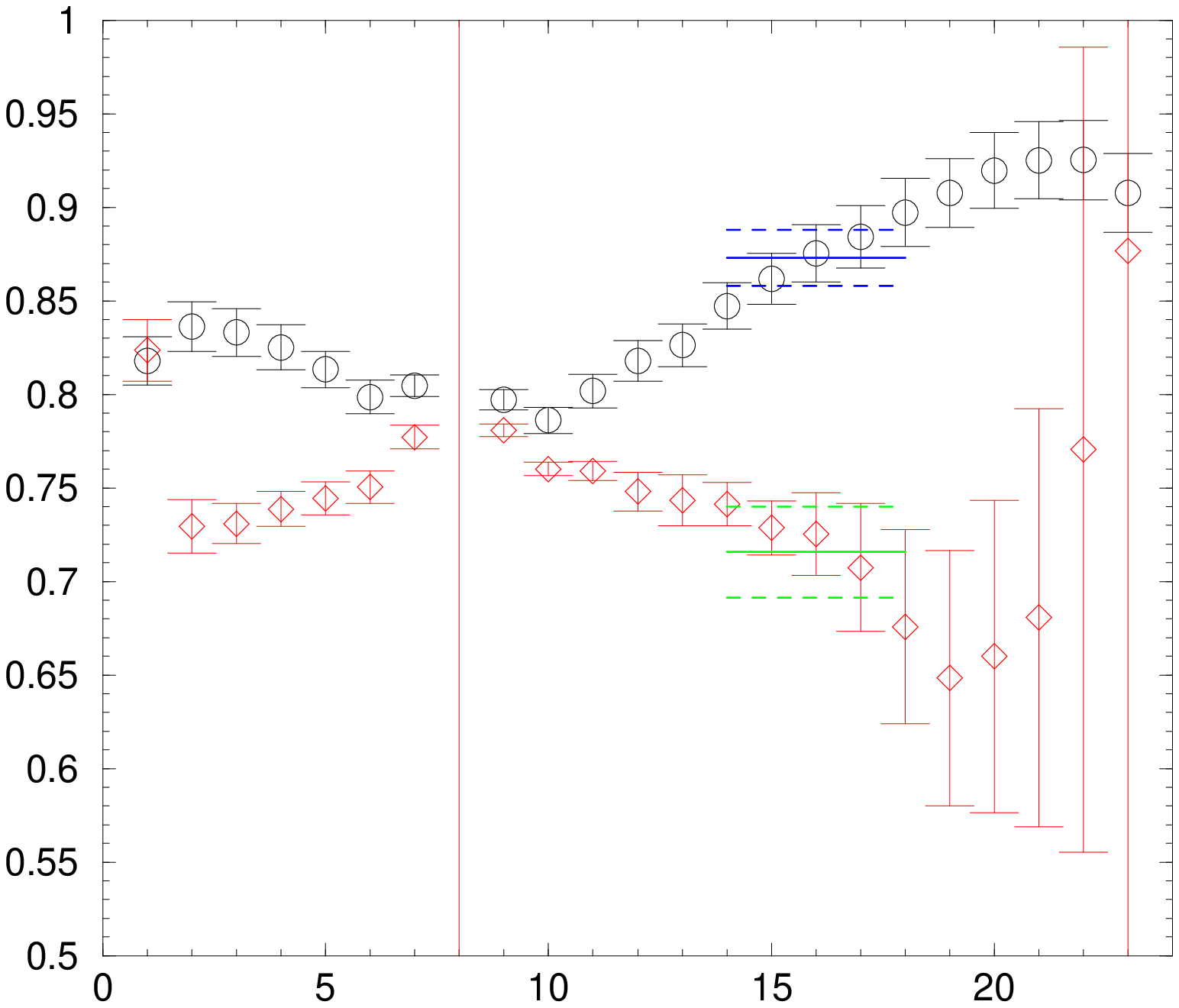}
  \includegraphics[width=5.cm]{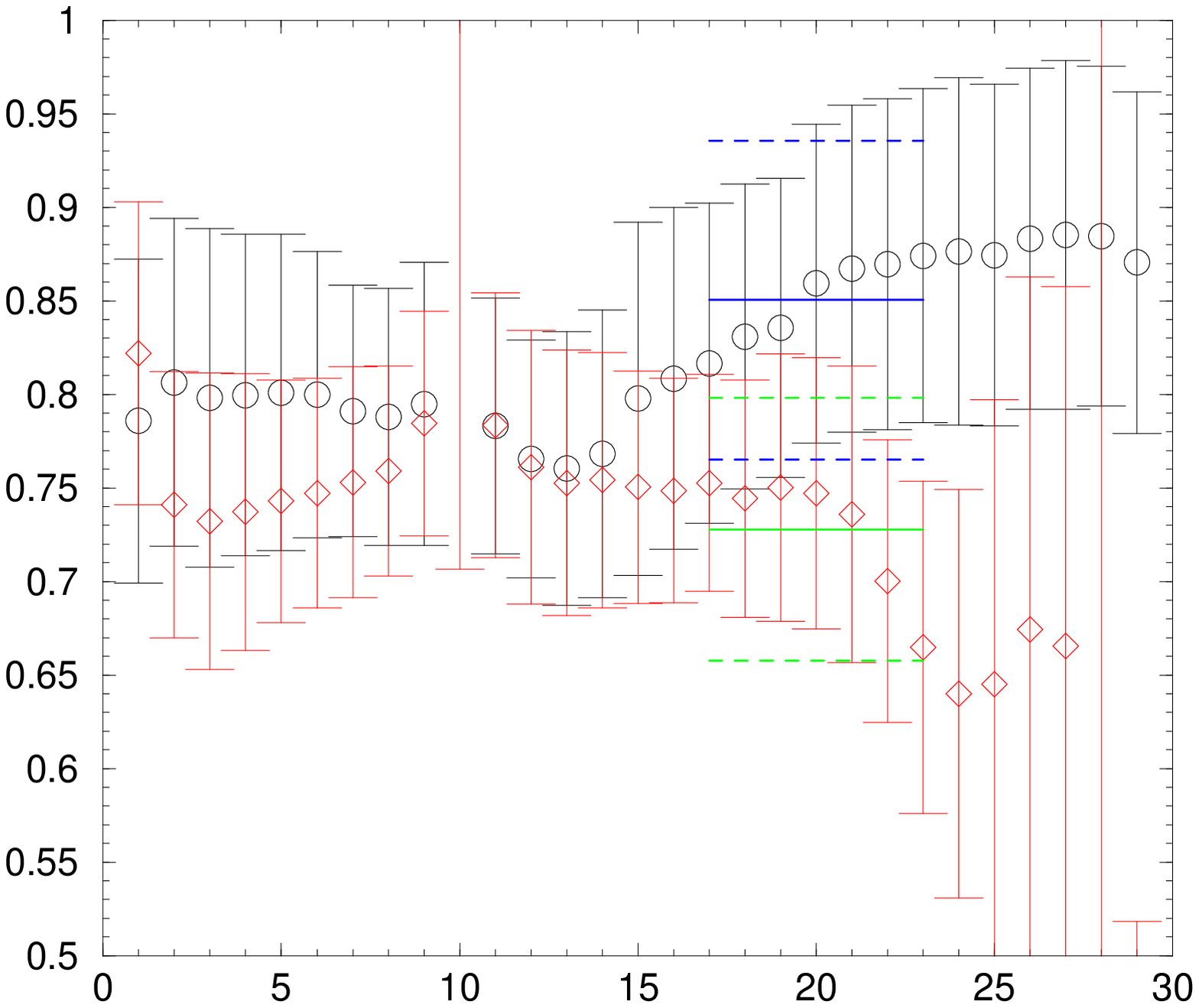}
  \caption{$x_0$ dependence of $\wt{Z}_A(g_0,m,x_0)$ at $\beta=1.90$ and
  $\theta=0.5$ for box sizes $8^3\times18$ (left),
  $10^3\times24$ (middle) and $12^3\times30$ (right).
  Red diamonds and black circles are definition with or without the
  disconnected diagrams.
  Solid and dashed lines represent an expected plateau, which is given
  by fitting data around $x_0=2T/3$ by a constant.
  Error is estimated with the Jackknife method.
}
 \label{fig:za-b190-th05}
 \end{center}
\end{figure}

\begin{figure}
 \begin{center}
  \includegraphics[width=5.cm]{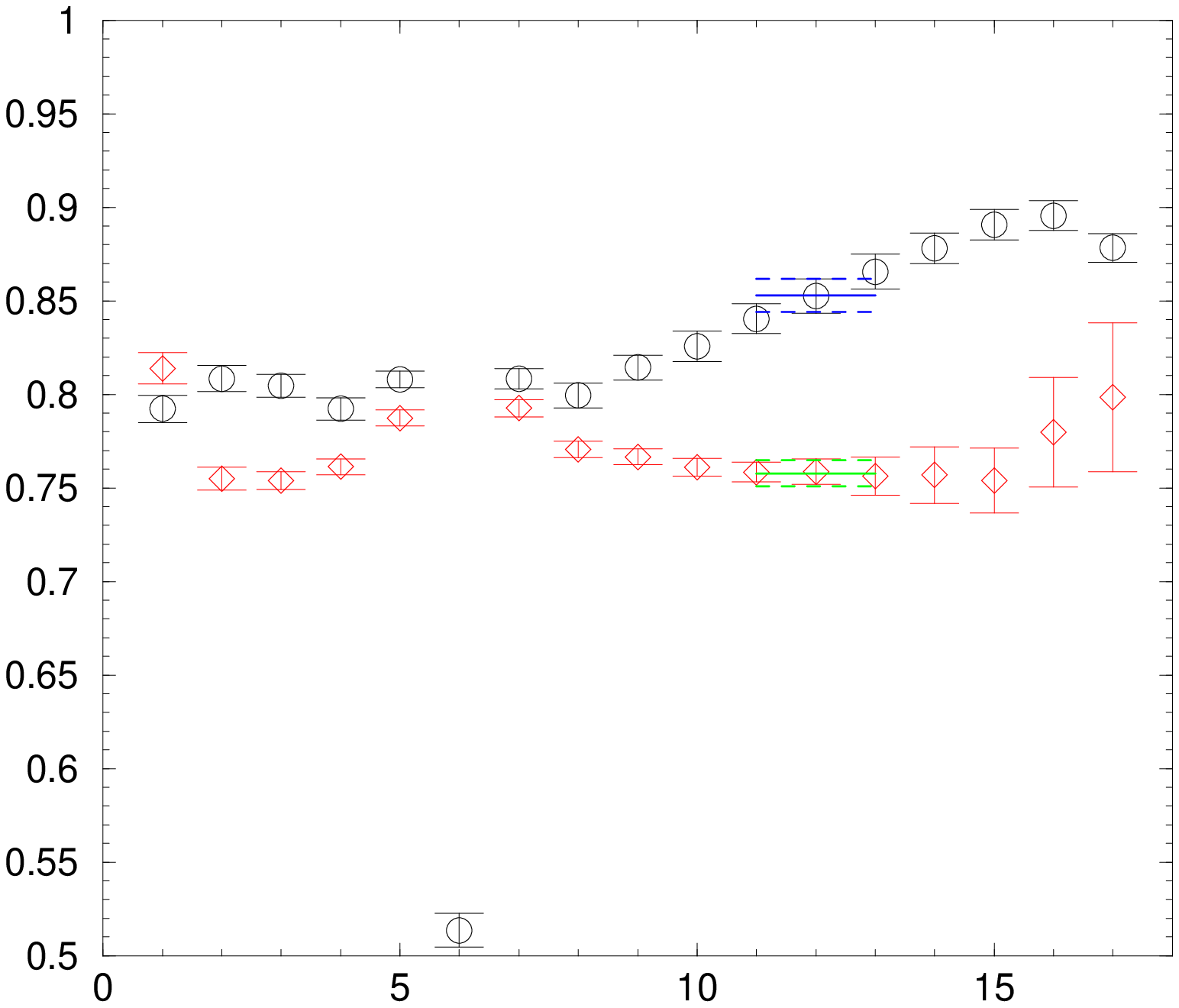}
  \includegraphics[width=5.cm]{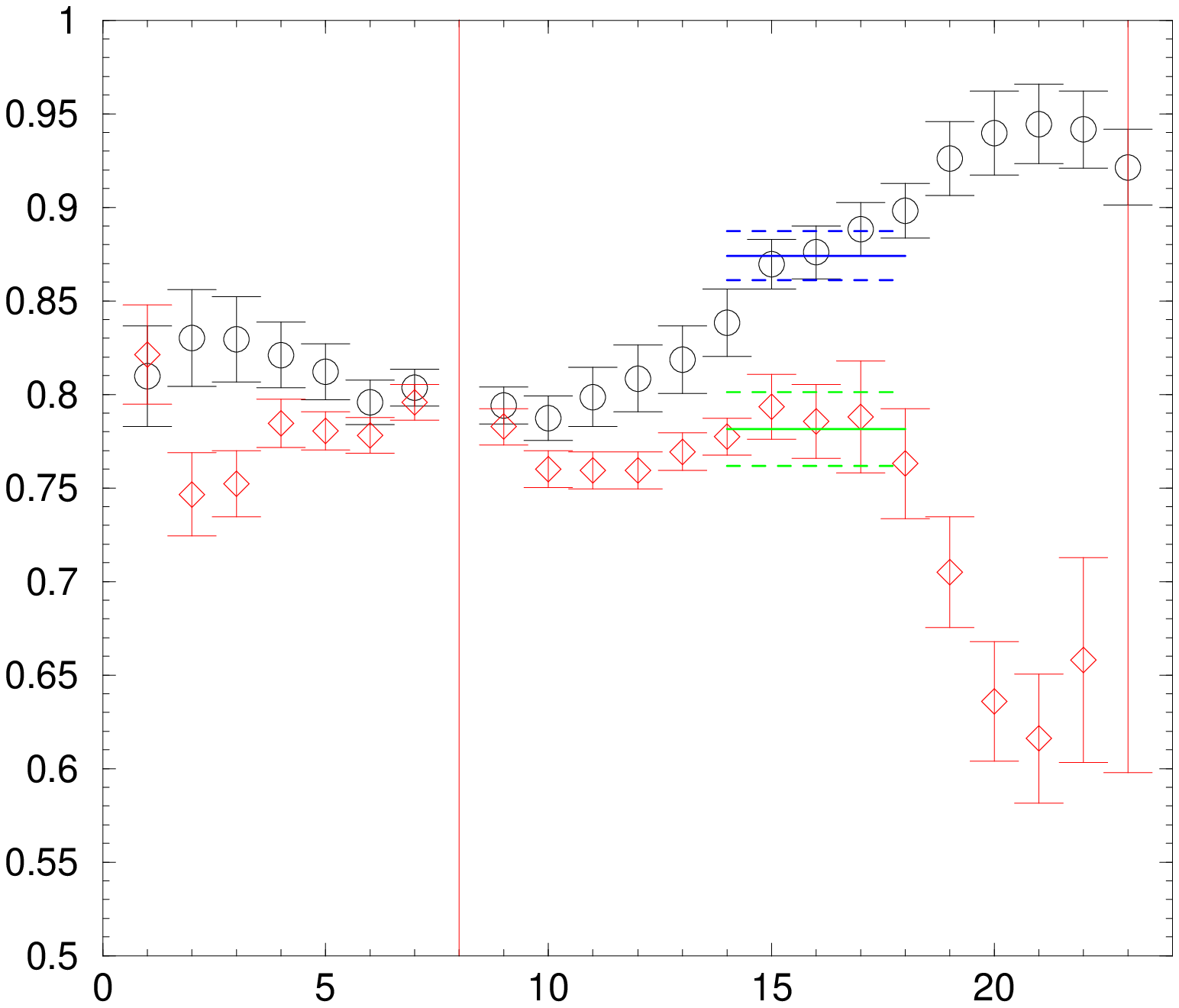}
  \caption{$x_0$ dependence of $\wt{Z}_A(g_0,m,x_0)$ at $\beta=1.90$ and
  $\theta=0$ for box sizes $8^3\times18$ (left) and $10^3\times24$ (right).}
 \label{fig:za-b190-th00}
 \end{center}
\end{figure}

\begin{figure}
 \begin{center}
  \includegraphics[width=5.cm]{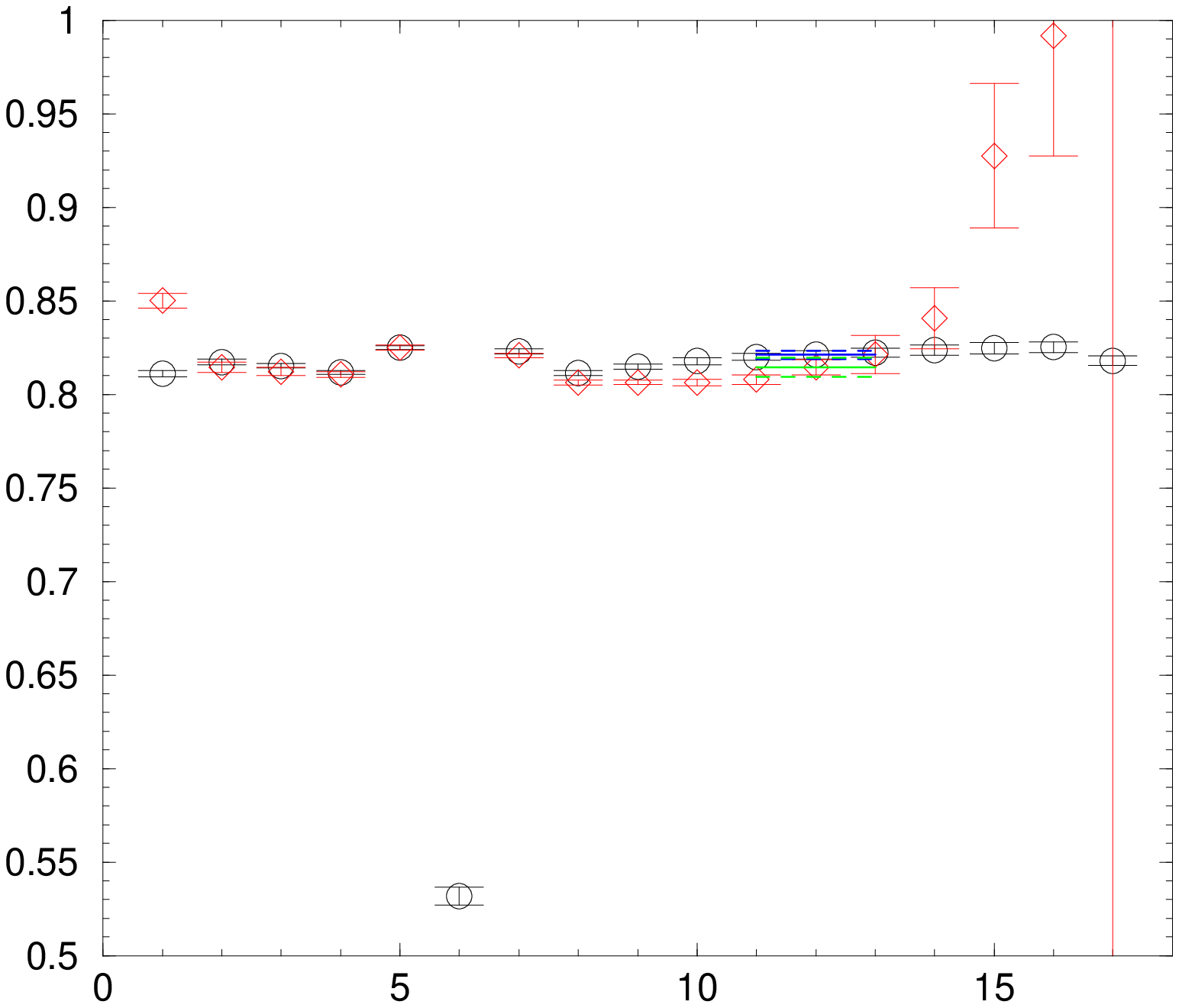}
  \includegraphics[width=5.cm]{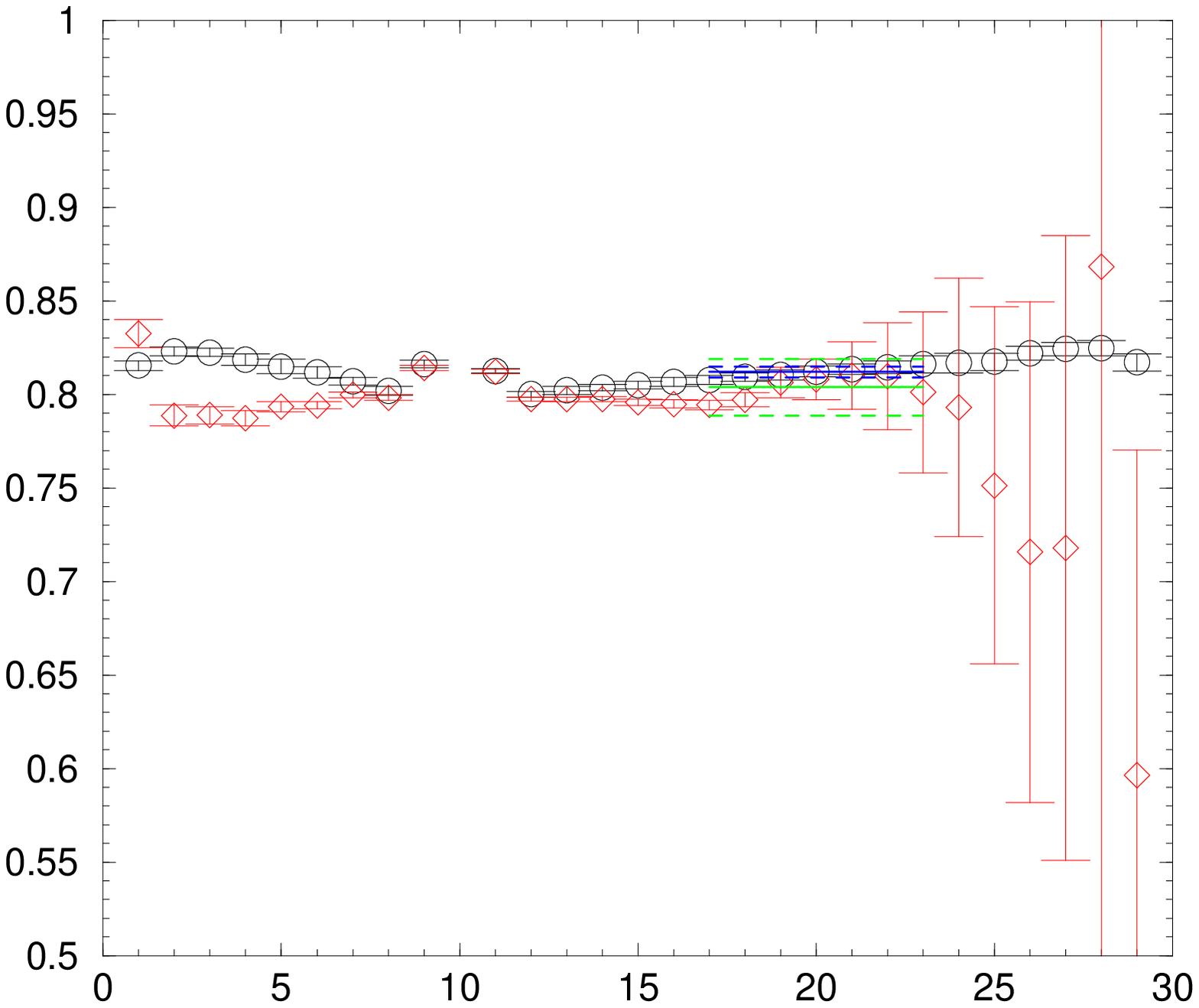}
  \caption{$x_0$ dependence of $\wt{Z}_A(g_0,m,x_0)$ at $\beta=2.05$ and
  $\theta=0.5$ for box sizes $8^3\times18$ (left) and $12^3\times30$ (right).}
 \label{fig:za-b205}
 \end{center}
\end{figure}

\begin{figure}
 \begin{center}
  \includegraphics[width=5.cm]{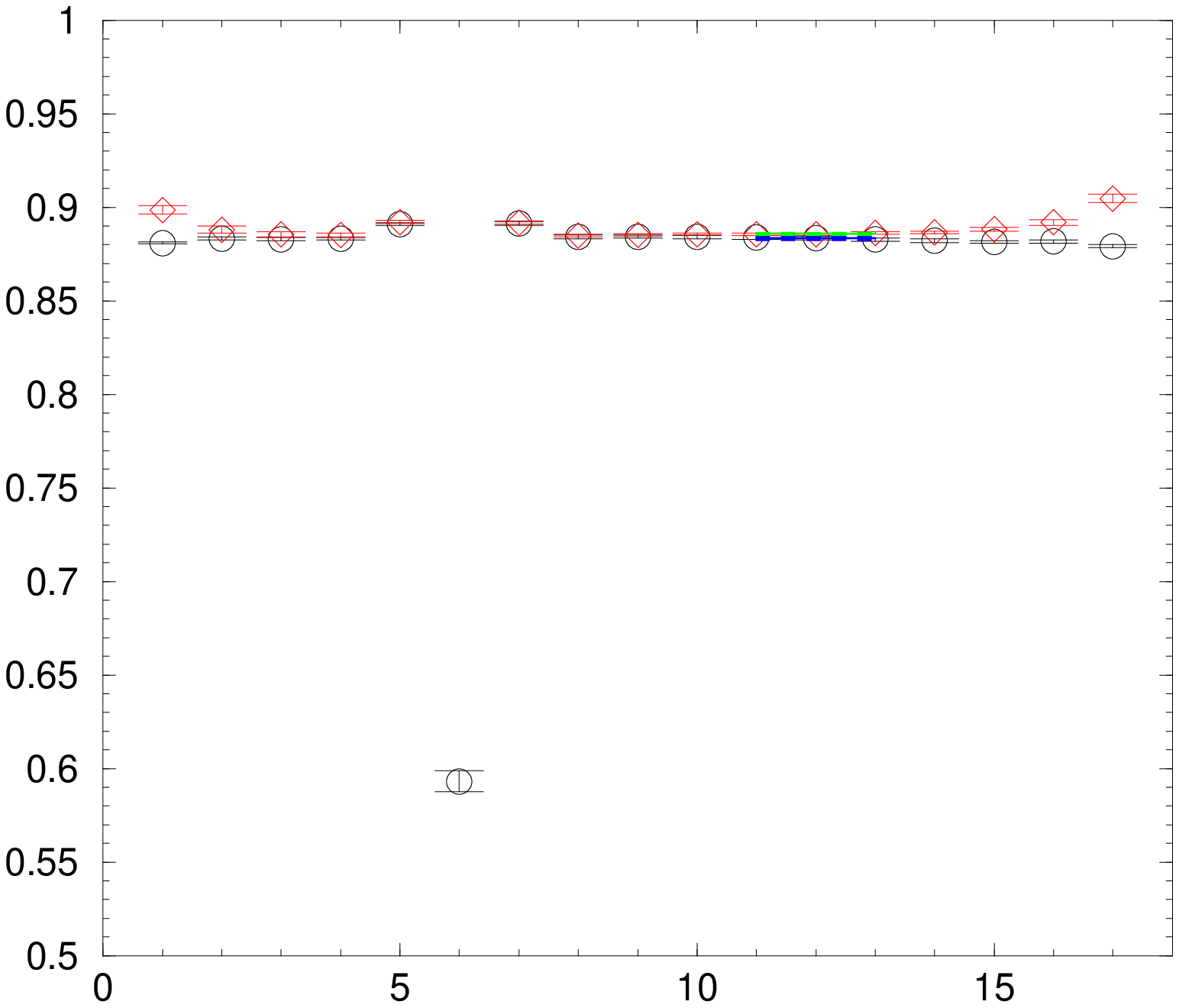}
  \includegraphics[width=5.cm]{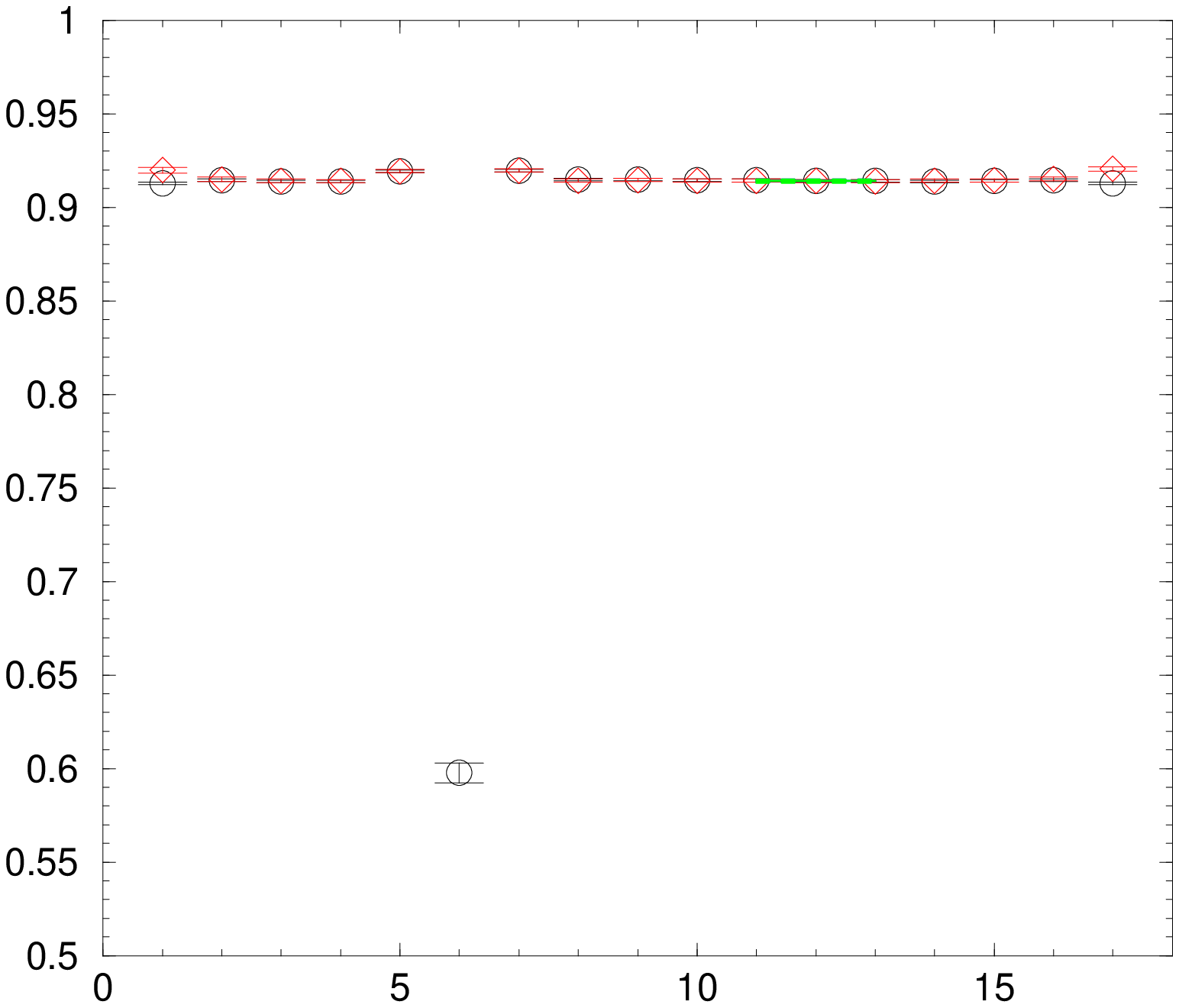}
  \includegraphics[width=5.cm]{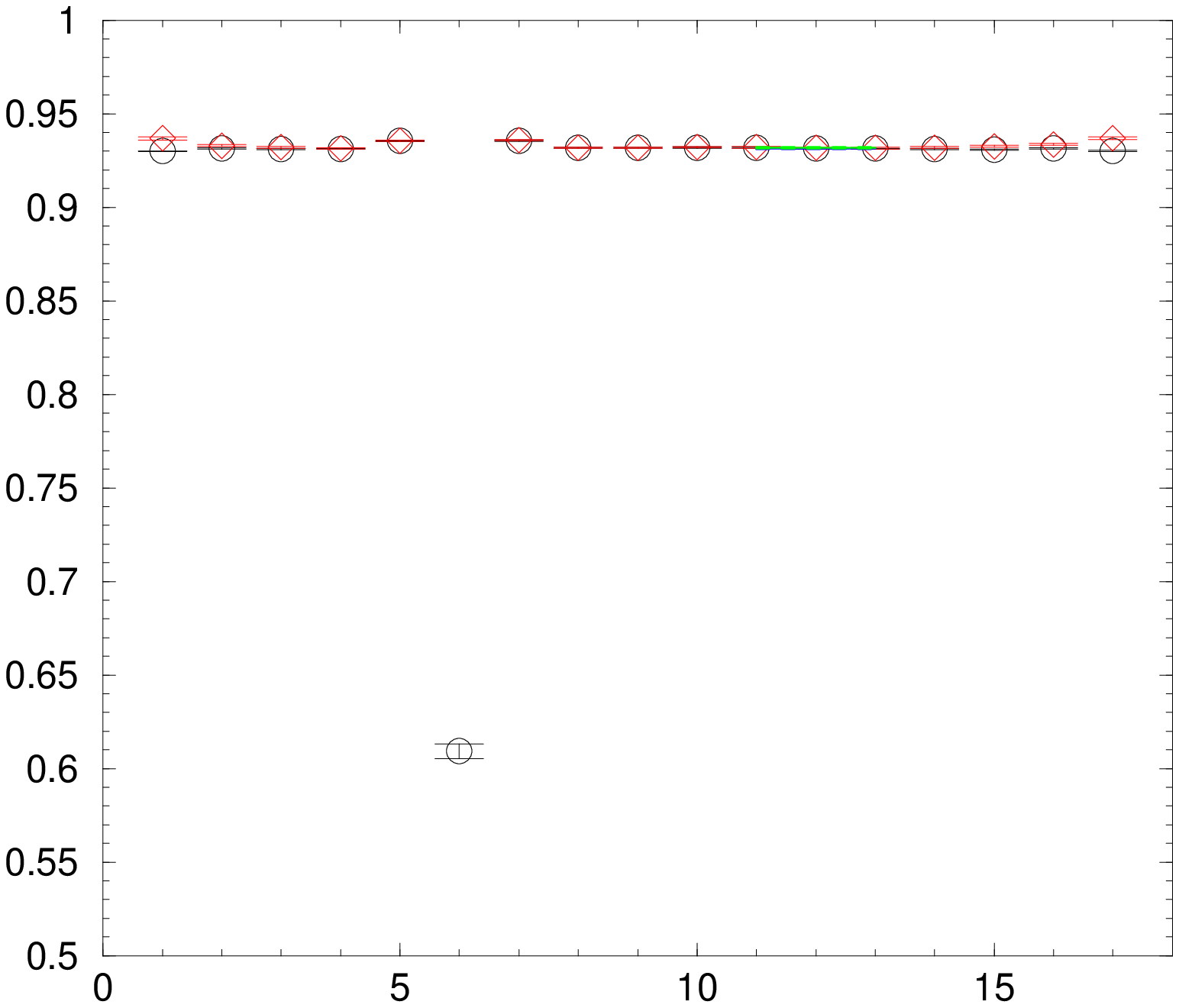}
  \caption{$x_0$ dependence of $\wt{Z}_A(g_0,m,x_0)$ at $\beta=3.0$
  (left), $4.0$ (middle) and $5.0$ (right).
  The box size is $8^3\times18$ and twist angle is set to $\theta=0.5$.}
 \label{fig:za-b30}
 \end{center}
\end{figure}

The time dependence is more tamed at $\beta=1.90$ as is seen in
Figs.~\ref{fig:za-b190-th05} and \ref{fig:za-b190-th00}.
We found a plateau in $\wt{Z}_A$ with the disconnected diagram for
$L/a\ge10$ at $\theta=0.5$ or $L/a\ge8$ with $\theta=0$.
The value of $Z_A$ is consistent with each other at $x_0=2T/3$ within
one standard deviation once a plateau is located.
We have no trouble to find a plateau for $\beta\ge2.05$ as is shown in
Figs.~\ref{fig:za-b205}, \ref{fig:za-b30}.
Furthermore results from the two definitions with or without the
disconnected diagrams are consistent with each other.

We conclude that we should adopt the definition with the
disconnected diagrams \cite{Della Morte:2005rd} to locate a
plateau at smaller $\beta$.
It is better to set the twist angle $\theta$ to zero although
computational cost is slightly heavier and statistical fluctuation
becomes worse.
The box size should be taken as large as possible
% unless we find a plateau 
since lattice artifacts should manifest as a polynomial of $a/L$.
%However it seems that we could not find a good plateau at $\beta=1.83$.
%Our definition of $Z_A$ may contain large lattice artifact at
%$\beta=1.83$.

In this paper we adopted the following box size at each $\beta$ to
define $Z_A$:
$12^3\times30$, $\theta=0.5$ at $\beta=1.83$, $10^3\times24$, $\theta=0$
at $\beta=1.90$ and $12^3\times30$, $\theta=0.5$ for $\beta=2.05$.
The disconnected contribution is included.
The physical box size varies from $0.84$ fm to $1.3$ fm.
We derive our $Z_A$ by fitting the plateau around $x_0=2T/3$ by a
constant.
The fitting range is plotted in
Figs.~\ref{fig:za-b183-th05} - \ref{fig:za-b30}.
The error is estimated by the Jackknife method.
The results are listed in tables \ref{tab:za-b183}, \ref{tab:za-b190}
and \ref{tab:za-b205} together with those at various box sizes.

\begin{figure}
 \begin{center}
  \includegraphics[width=8.0cm]{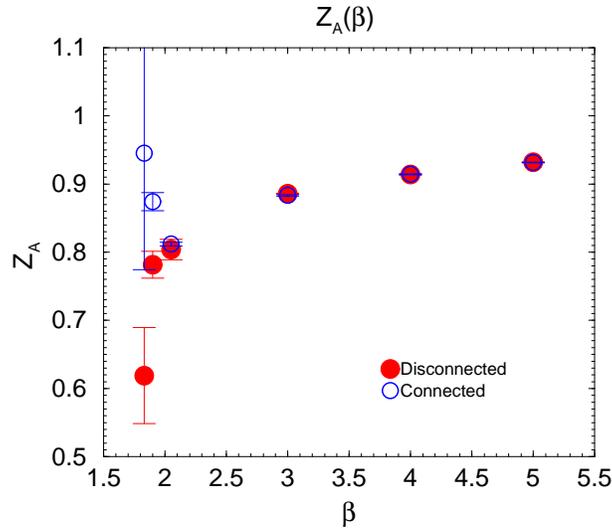}
  \caption{$\beta$ dependence of $Z_A(g_0)$ with disconnected diagrams
  (filled circles) and that without them (open circles).}
  \label{fig:za-beta}
 \end{center}
\end{figure}

In Fig.~\ref{fig:za-beta} we plot the values for $Z_A$ according to our
definition (filled circles), together with those of other definitions
(open symbols).
Scattering of points starting around $\beta=1.95$ indicates that lattice
artifacts are increasingly large in our data for large lattice spacings.

\begin{figure}
 \begin{center}
  \includegraphics[width=5.5cm]{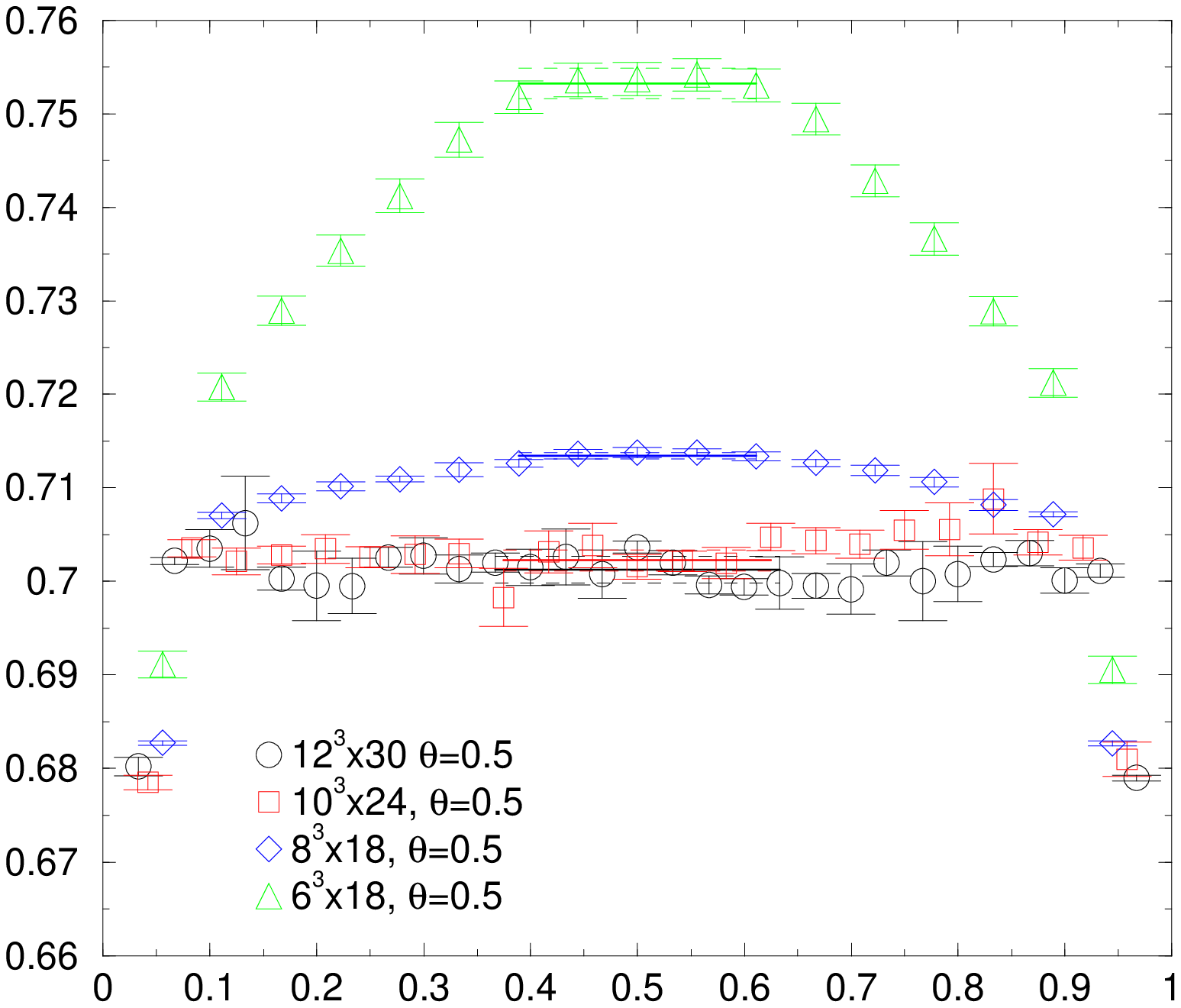}
  \includegraphics[width=5.5cm]{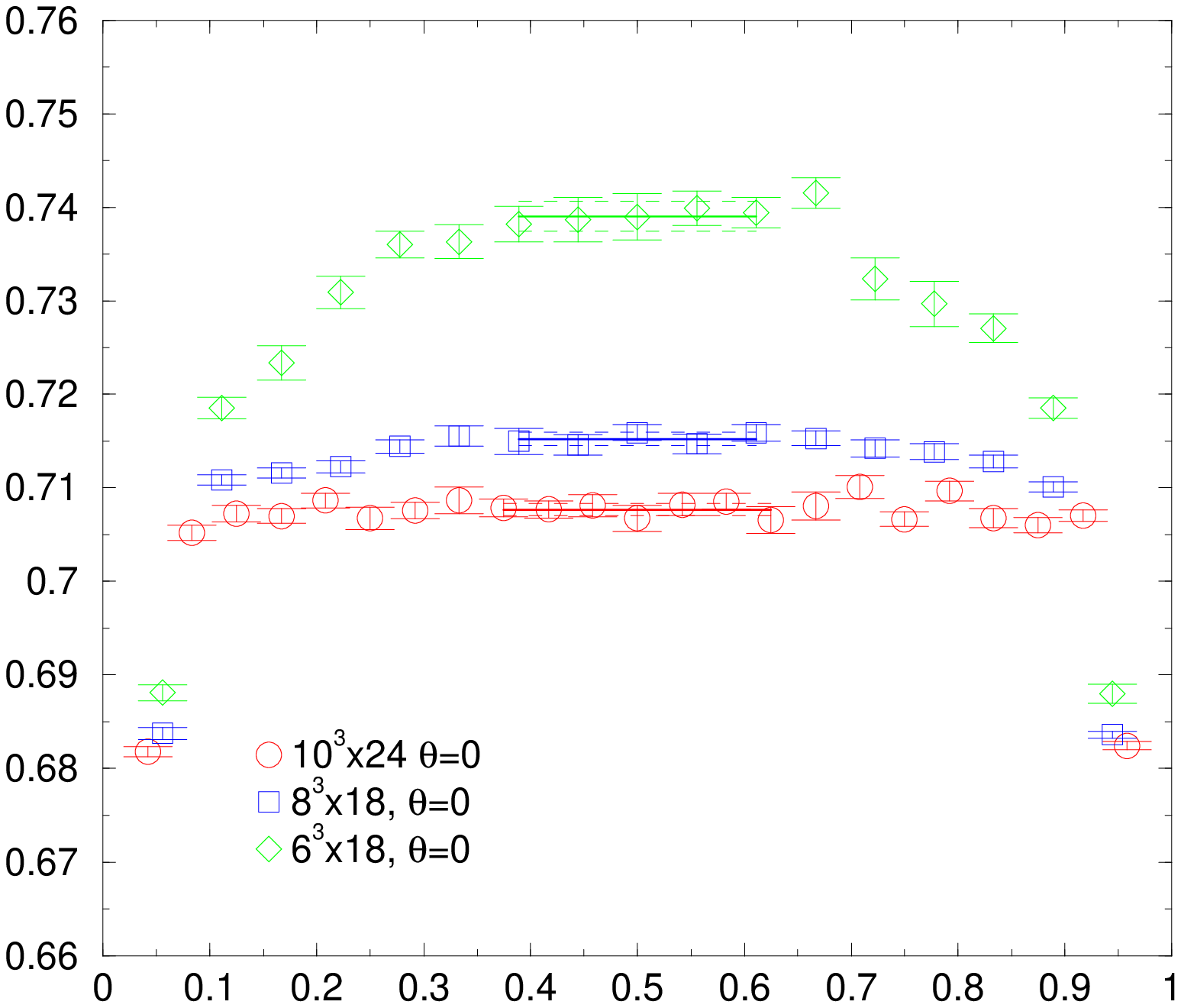}
  \caption{$x_0$ dependence of $\wt{Z}_V(g_0,x_0)$ at $\beta=1.83$,
  $\theta=0.5$ (left) and $\theta=0$ (right)
  for box sizes $6^3\times18$ (green triangles),
  $8^3\times18$ (blue diamonds), $10^3\times24$ (red squares) and
  $12^3\times30$ (black circles).
  Solid and dashed lines represent an expected plateau, which is given
  by fitting data around $x_0=T/2$ by a constant.
}
 \label{fig:zv-b183}
 \end{center}
\end{figure}

\begin{figure}
 \begin{center}
  \includegraphics[width=5.cm]{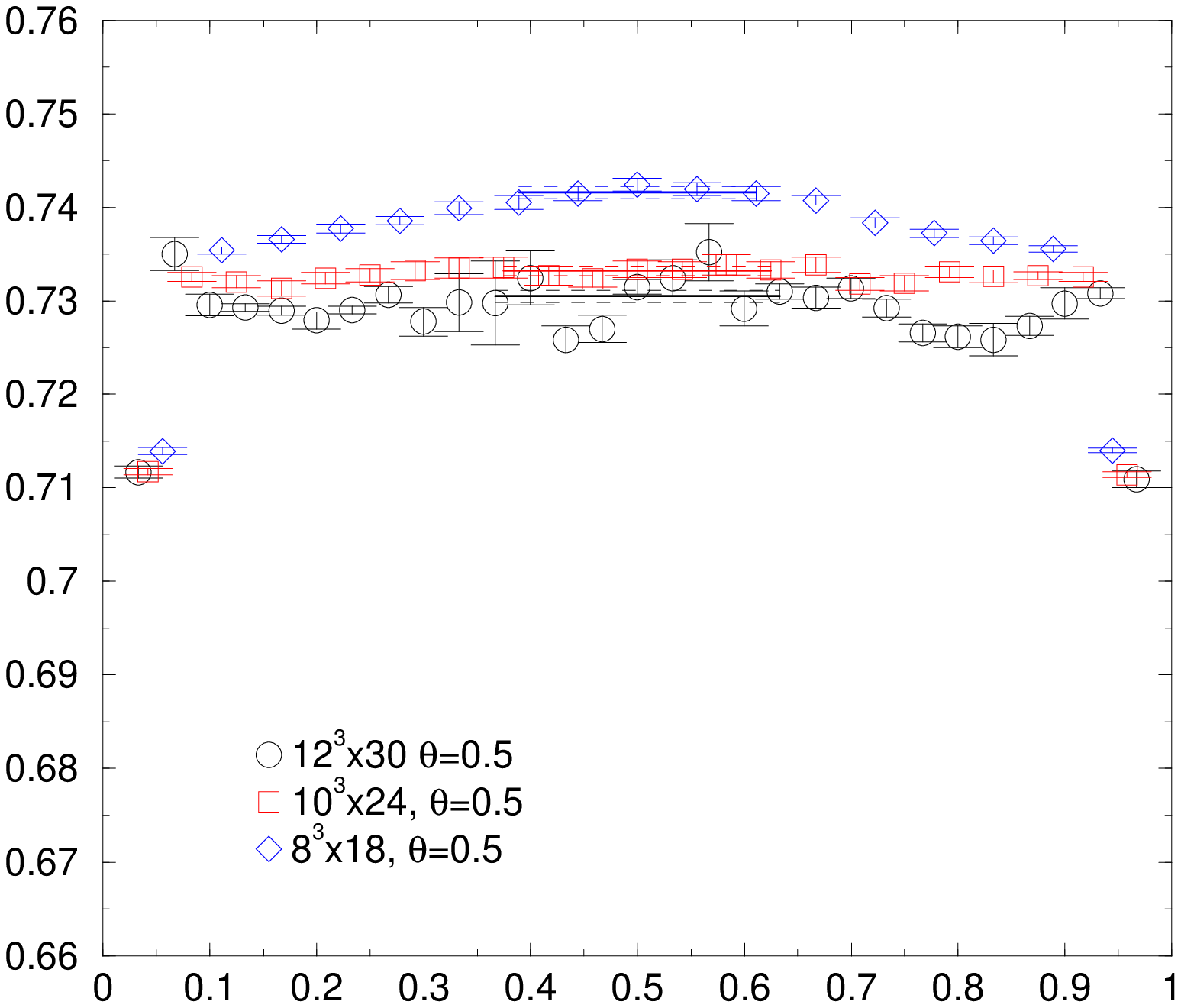}
  \includegraphics[width=5.cm]{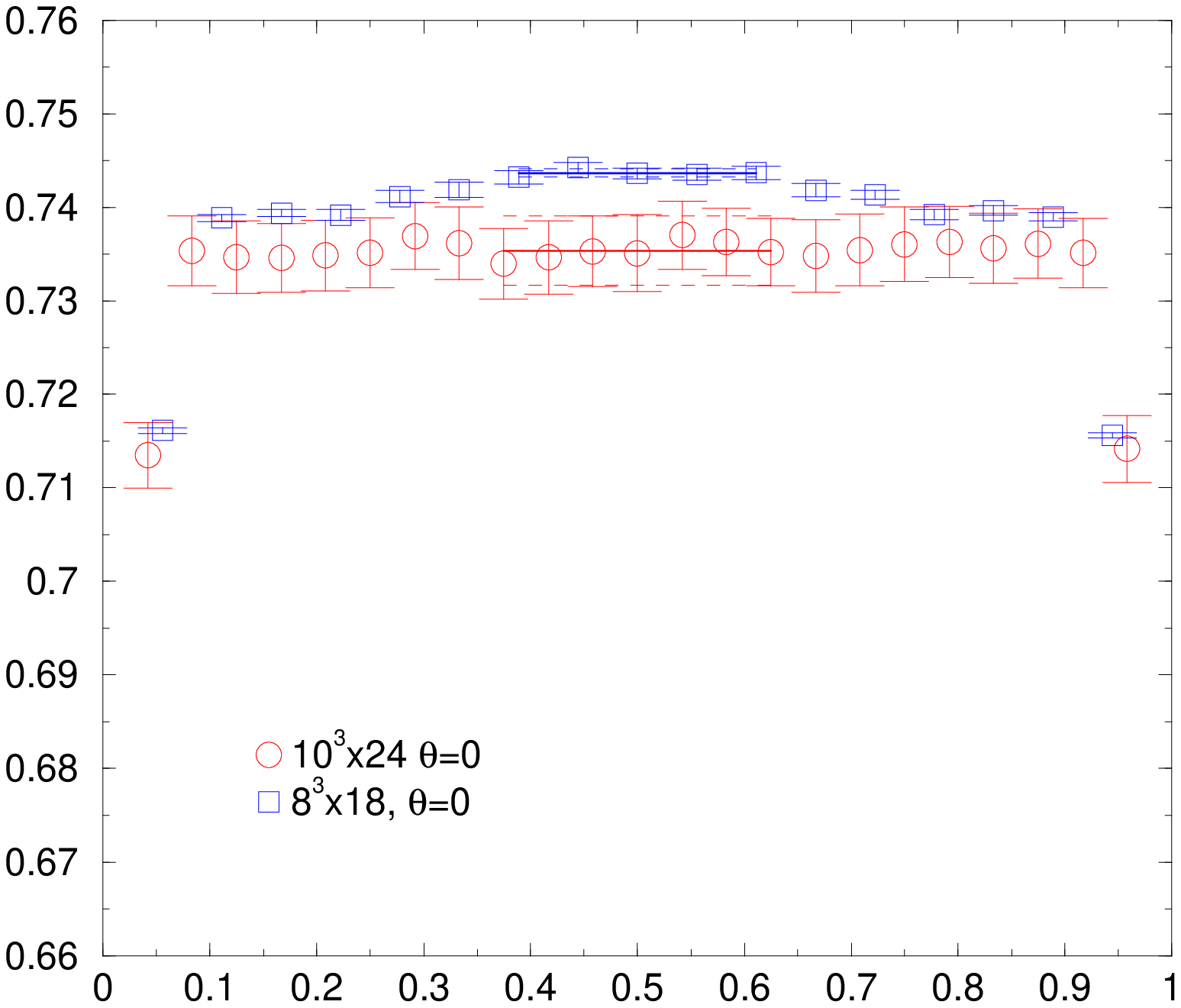}
  \caption{$x_0$ dependence of $\wt{Z}_V(g_0,x_0)$ at $\beta=1.90$,
  $\theta=0.5$ (left) and $\theta=0$ (right)
  for box sizes $8^3\times18$ (blue diamonds),
  $10^3\times24$ (red squares) and $12^3\times30$ (black circles).}
 \label{fig:zv-b190}
 \end{center}
\end{figure}

\begin{figure}
 \begin{center}
  \includegraphics[width=5.cm]{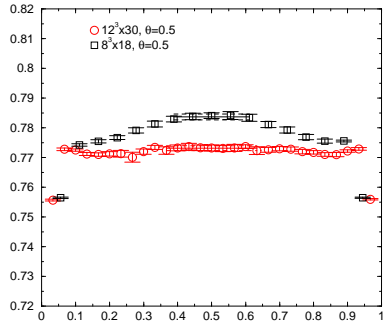}
  \caption{$x_0$ dependence of $\wt{Z}_V(g_0,m,x_0)$ at $\beta=2.05$ and
  $\theta=0.5$ for box sizes $8^3\times18$ (black squares) and
  $12^3\times30$ (red circles).}
 \label{fig:zv-b205}
 \end{center}
\end{figure}

\begin{figure}
 \begin{center}
  \includegraphics[width=5.cm]{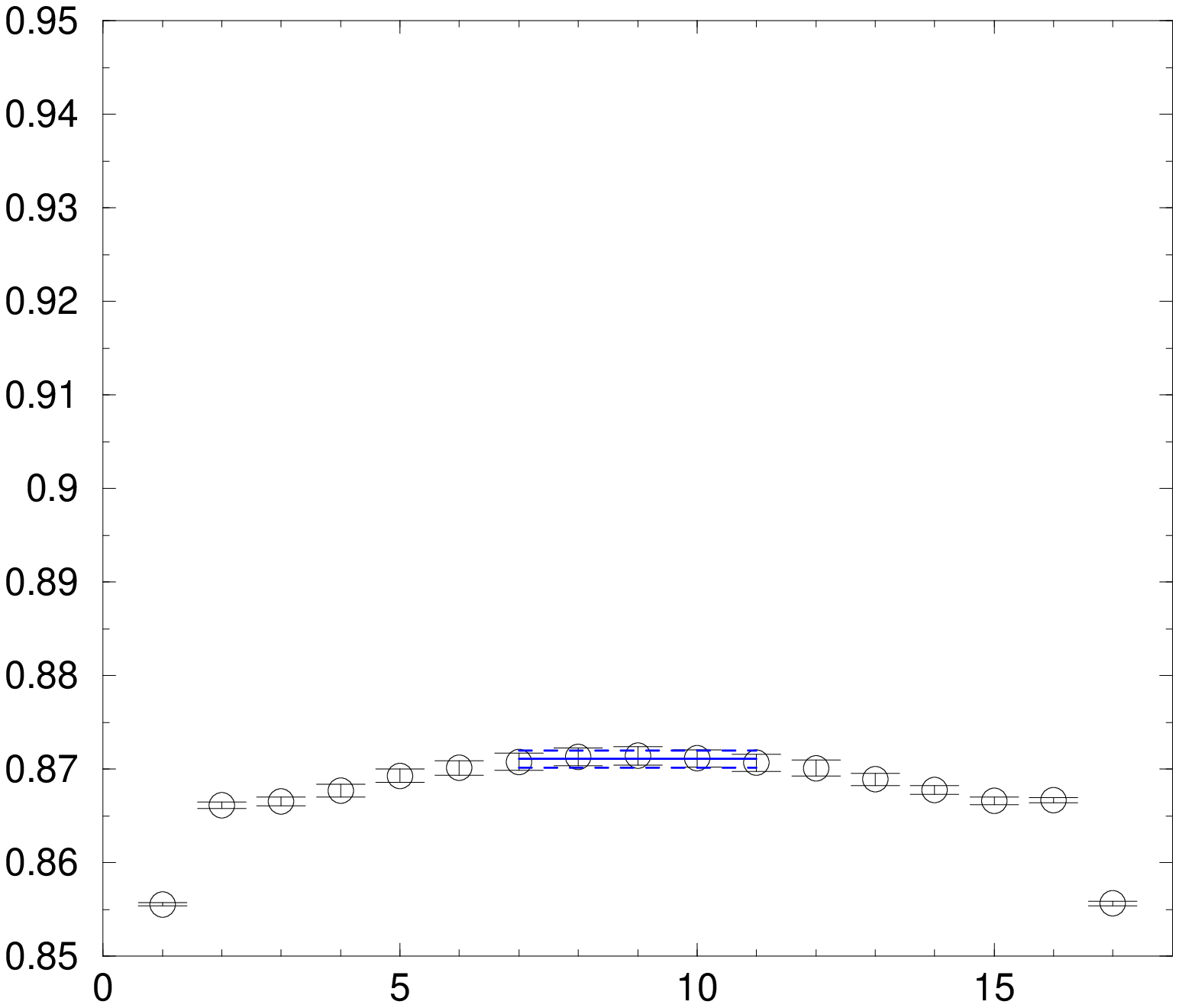}
  \includegraphics[width=5.cm]{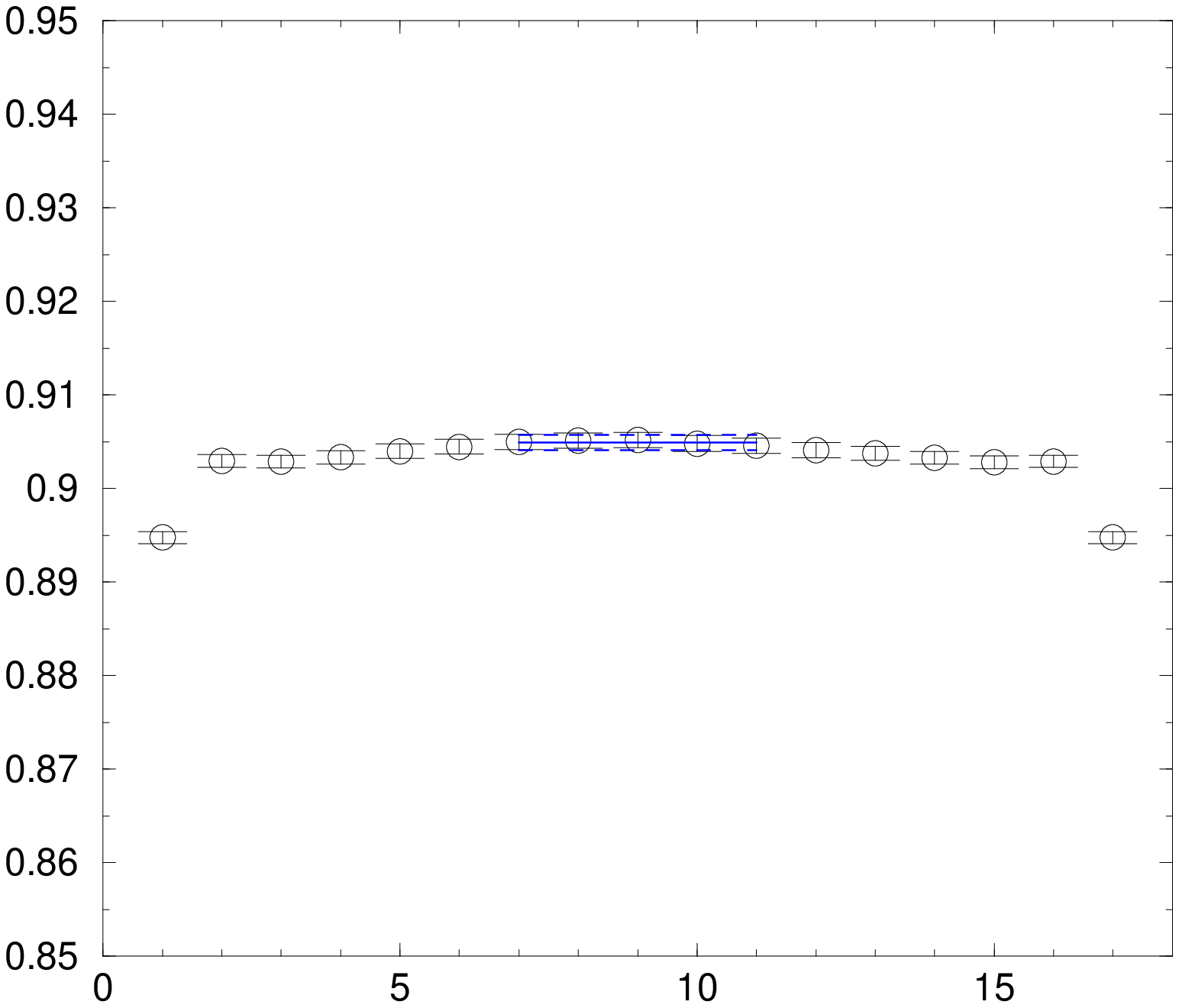}
  \includegraphics[width=5.cm]{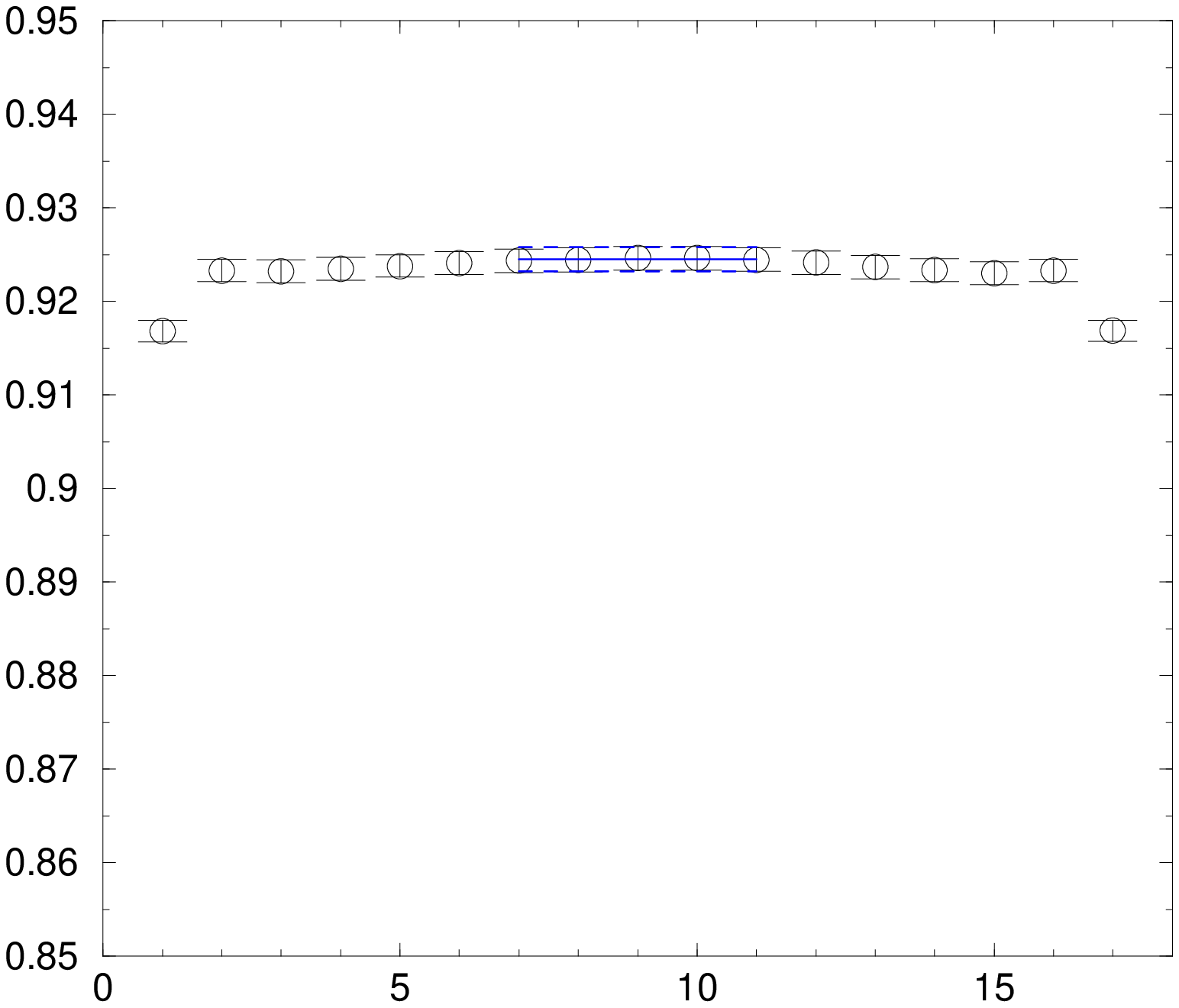}
  \caption{$x_0$ dependence of $\wt{Z}_V(g_0,x_0)$ at $\beta=3.0$
  (left), $\beta=4.0$ (middle) and $\beta=5.0$ (right).
  A box size $8^3\times18$ and a value of $\theta=0.5$ is
  adopted.}
 \label{fig:zv-b30}
 \end{center}
\end{figure}

The time dependence of $\wt{Z}_V(g_0,x_0)$ is plotted for $\beta=1.83$
in Fig.~\ref{fig:zv-b183} for $\theta=0.5$ (left) and $\theta=0$
(right) with horizontal axis normalized to unity.
For $L/a\ge10$ we find a long plateau whose values does not depend on
$L$ and $\theta$.
The behavior is almost the same for $\beta=1.90$
(Fig.~\ref{fig:zv-b190}) and $\beta=2.05$ (Fig.~\ref{fig:zv-b205}).
%where we find a good plateau for $L/a\ge10$, and 
$\wt{Z}_V(g_0,x_0)$ seems to be almost flat at $\beta\ge3.0$ for $L/a=8$
as is seen in Fig.~\ref{fig:zv-b30}.
%On the other hand,
%$Z_V$ turned out to have large box size dependence for small volume even if 
%we find a (short) plateau.
%We must use large enough box size to obtain reliable estimates.
The renormalization factor $Z_V$ given by fitting the plateau is listed
in tables \ref{tab:za-b183}, \ref{tab:za-b190} and \ref{tab:za-b205}.

\begin{figure}
 \begin{center}
  \includegraphics[width=8.0cm]{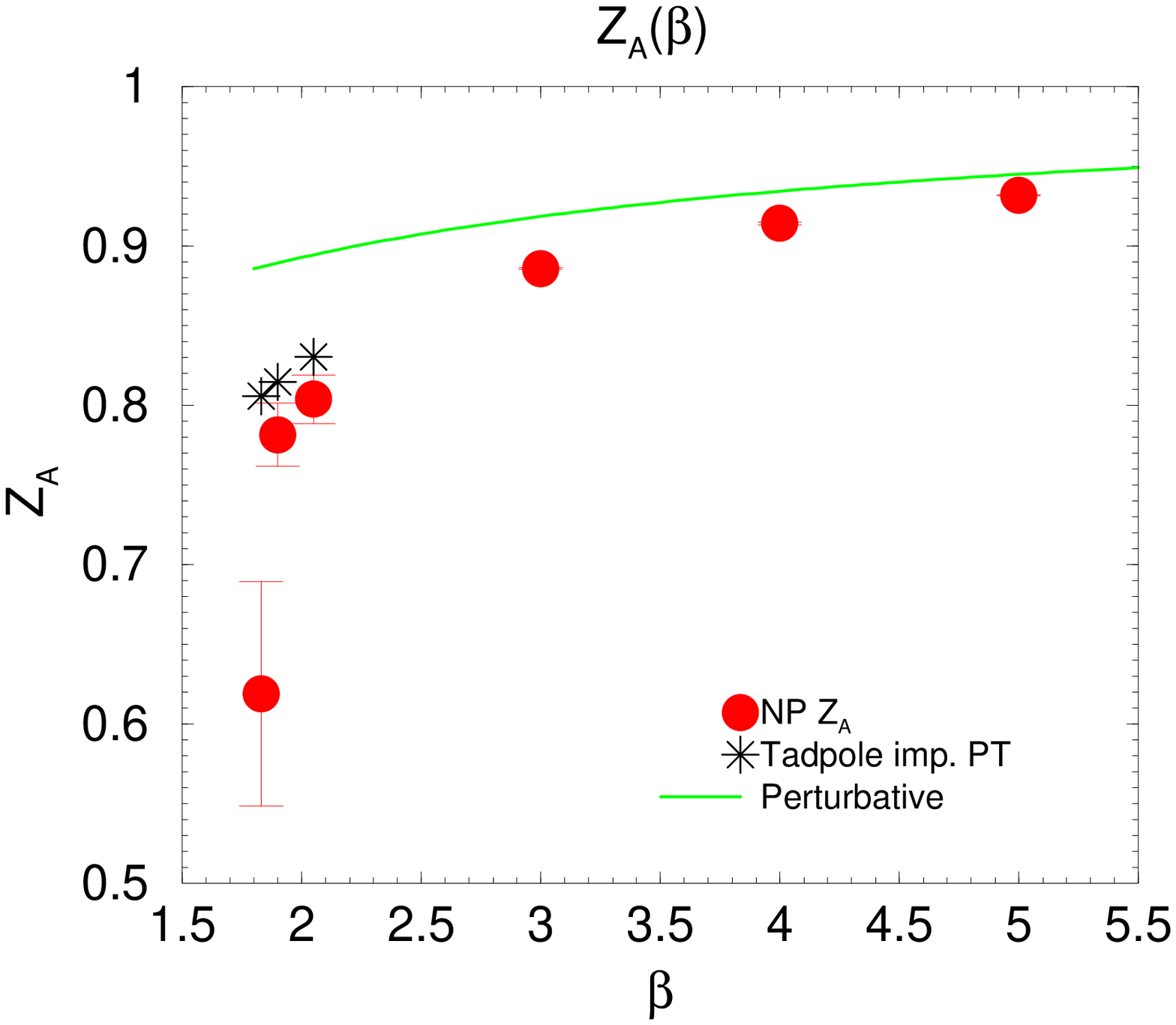}
  \includegraphics[width=8.0cm]{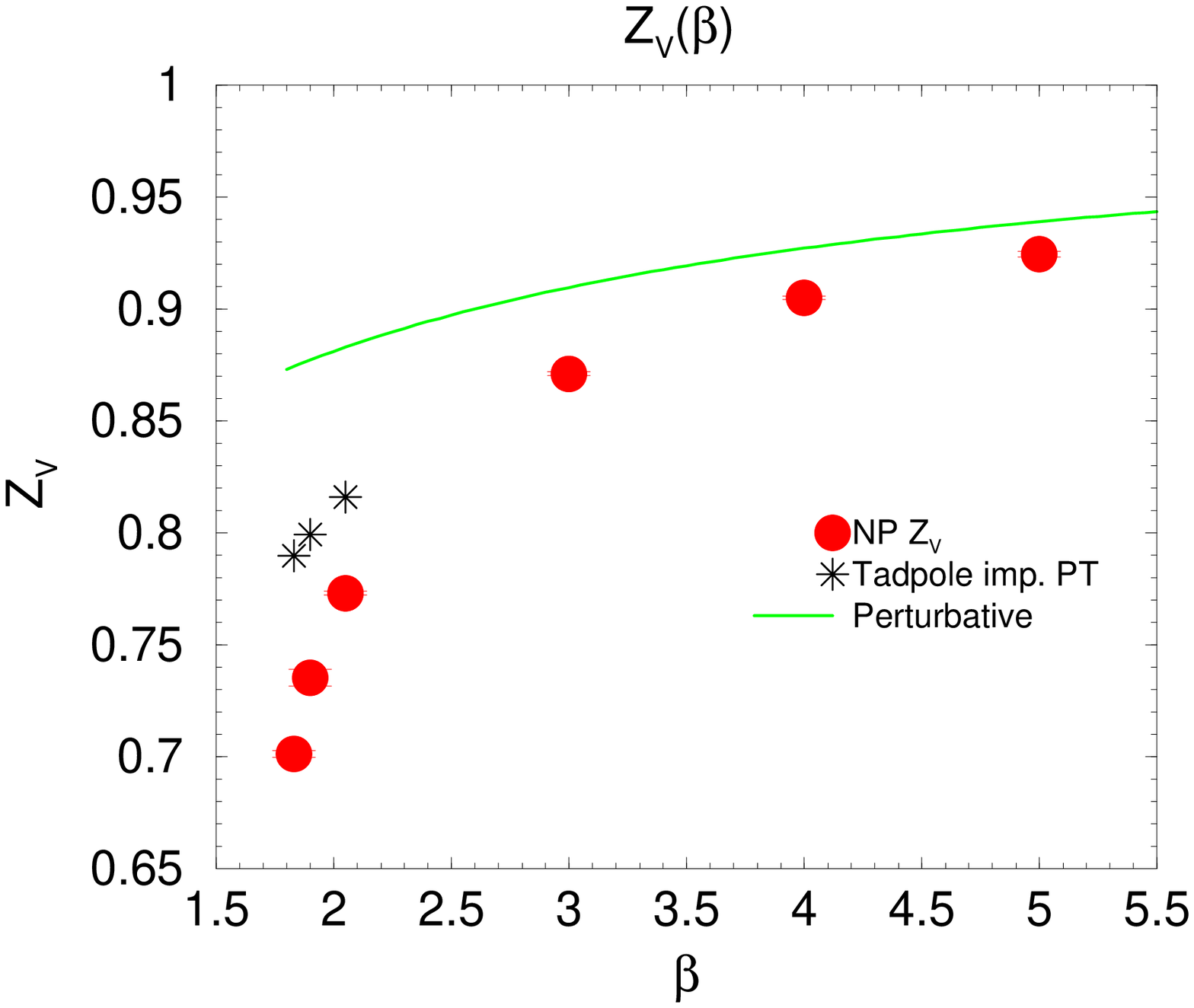}
  \caption{$\beta$ dependence of $Z_A(g_0)$ (left) and $Z_V(g_0)$ (right).
  Red filled circles are our non-perturbative renormalization factor.
  Solid line is a perturbative result at one loop.
  Stars are from tadpole improved perturbation theory.}
  \label{fig:zv-beta}
 \end{center}
\end{figure}

In Fig.~\ref{fig:zv-beta} we plot our final results for
$Z_A(g_0)$ and $Z_V(g_0)$ as a function of $\beta$ 
together with perturbative behavior (solid line) and
results from the tadpole improved perturbation theory (star symbols).
%$Z_A$ without disconnected diagrams behaves in a non-monotonic way as
%was reported in Ref.~\cite{Della Morte:2005rd}.

\reseteqnum
%=============================================
\section{RGI mass renormalization factor}
%=============================================

We derive the renormalization factor $Z_M$ for the RGI mass according to
the procedure given in Sec.~\ref{sec:strategy}.
This factor is intended to renormalize the bare PCAC masses obtained in
the two large scale simulations carried out at three values of $\beta$.
The three hadron masses, $m_\pi$, $m_K$, $m_\Omega$ were used in
Ref.~\cite{Aoki:2008sm,Kuramashi} to determine the light quark masses
and the lattice spacing.
Two choices $m_\pi$, $m_\rho$ $m_K$ or $m_\pi$, $m_\rho$ $m_\phi$ were
adopted in Ref.~\cite{Ishikawa:2007nn}.
%We use the lattice spacings in these simulations to give intermediate
%physical scale $L_{\rm max}$.
The results for $Z_M$ are listed in Table \ref{tab:zm}.
Also listed are the mass renormalization factors
$Z_m^{\ovl{\rm MS}}(\beta,\mu=2\ {\rm GeV})$ in the $\ovl{\rm MS}$
scheme at a renormalization scale $\mu=2$ GeV.
We emphasize that the renormalization factor here is defined in terms
of the renormalization group functions for three flavors.
Results from tadpole improved perturbation theory is also listed for
comparison.

The error in the renormalization factor includes all the statistical
and systematic ones except for that from the choice of the reference
scale $L_{\rm max}$.
Lattice artifacts may be present at order ${\cal O}((a/L_{\rm max})^2)$.
In order to estimate this effect we increase $L_{\rm max}/a$ at each
$\beta$.
There is a problem that values of the renormalized coupling
$u_{\rm max}$ and the renormalization factor $Z_P(L_{\rm max})$ tend to
exceed the region covered by our step scaling function as is shown in
Table \ref{tab:Lmax}.
Since we attempt just a rough estimate, we adopted the step
scaling function by extrapolating its polynomial form.
The result is listed in Table \ref{tab:zmLmax} and
by comparing with values in Table \ref{tab:zm} we find a few percent
effect at $\beta=2.05$, while the magnitude may increase to a 10 \%
level at lower values of $\beta$.
%and we find the numerical vale of $Z_M$ is essentially consistent with
%that in the smaller box size within one standard deviation for
%$\beta=2.05$ and $1.83$.
%It seems that the lattice artifact due to a choice of
%$L_{\rm max}$ is not so severe for $\beta=2.05$.
%However we may have a systematic error of a few \% level at
%$\beta=1.90$, and we need to evaluate the step scaling function at
%stronger coupling to get more definite conclusion.

\begin{figure}[htbp]
\begin{center}
 \includegraphics[width=7.cm]{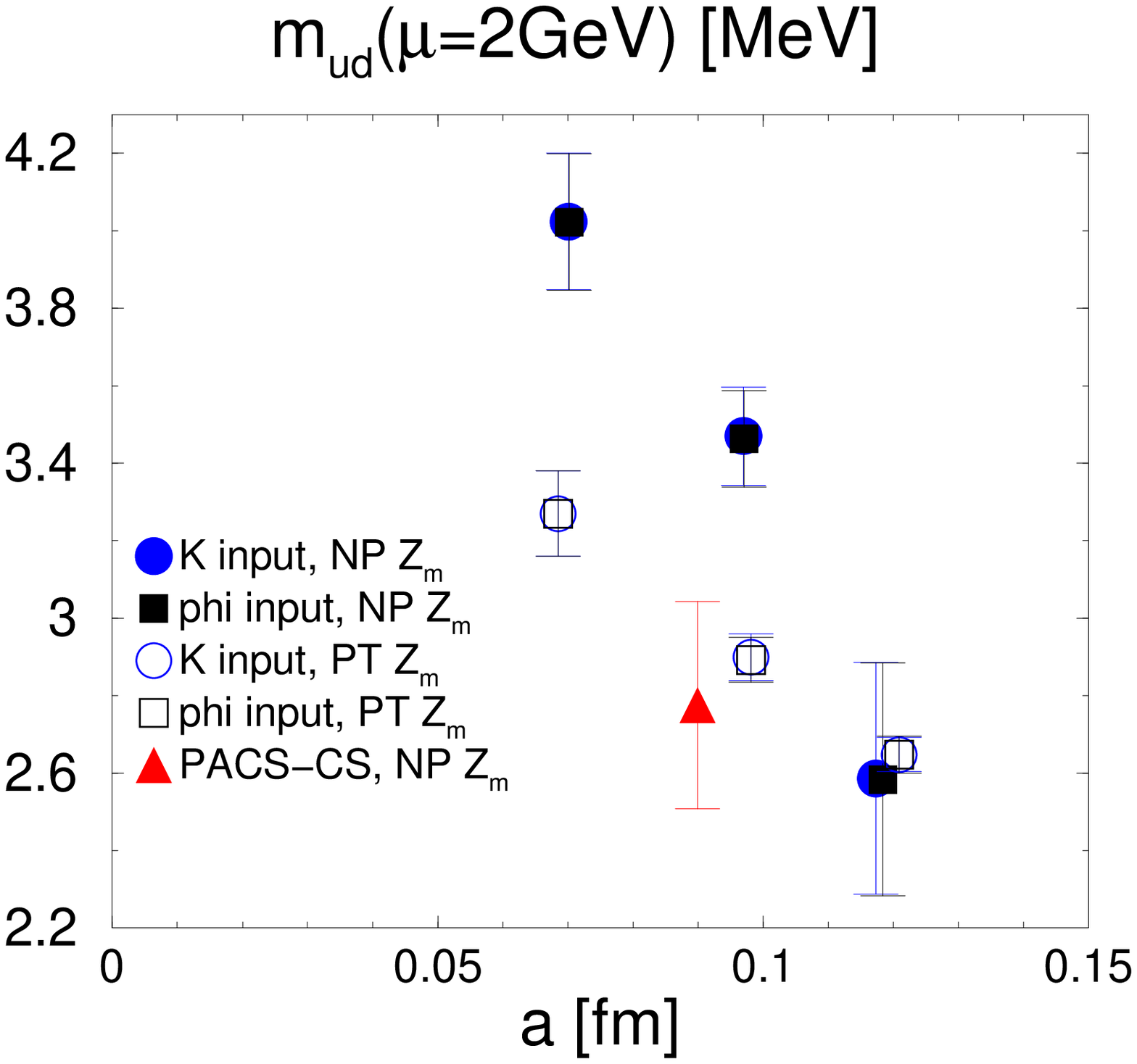}
 \includegraphics[width=7.cm]{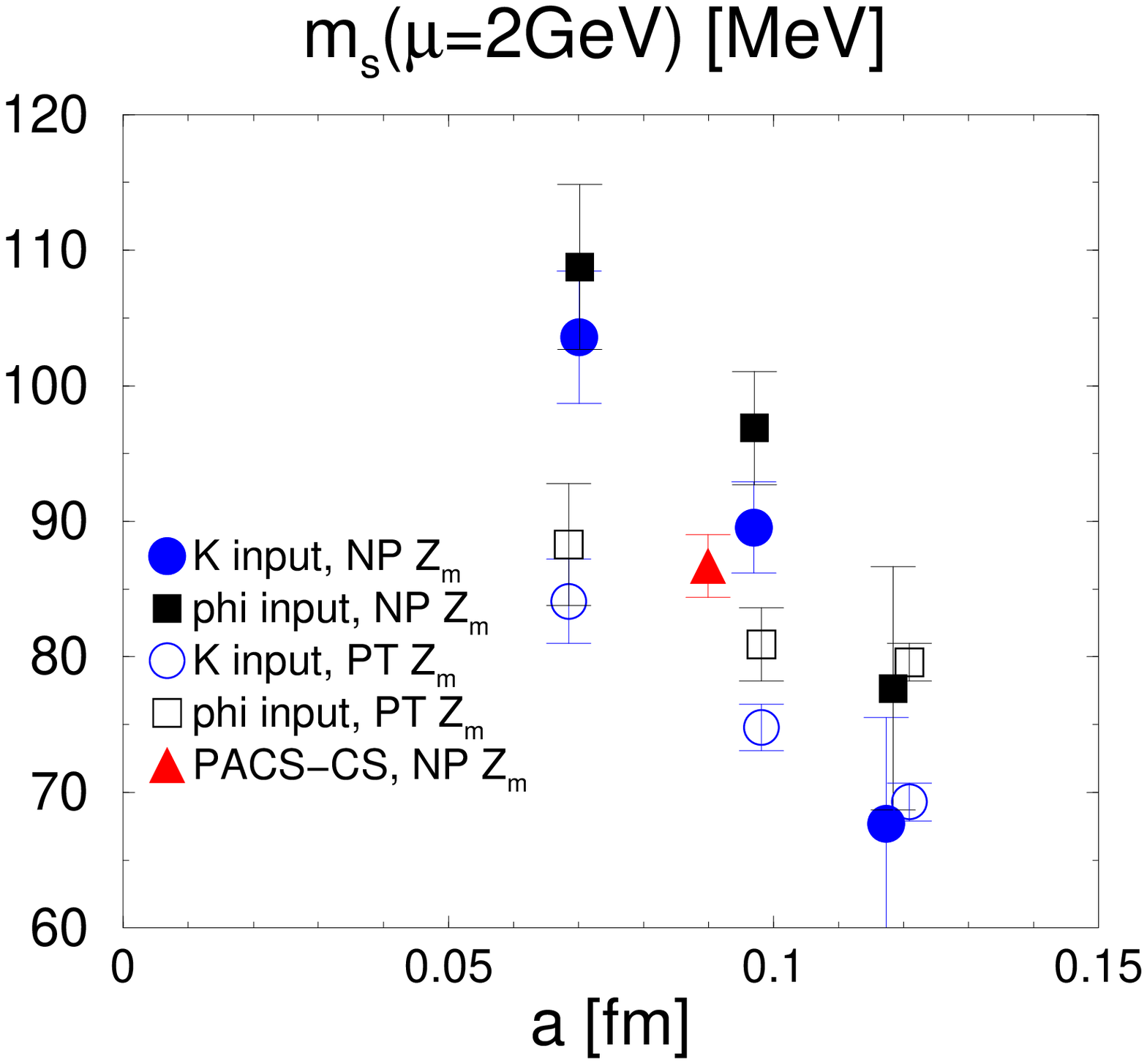}
 \caption{Scaling behavior of $m_{ud}^{\ovl{\rm MS}}$ (left) and
 $m_{s}^{\ovl{\rm MS}}$ (right); filled circles and
 squares are from CP-PACS/JLQCD result \cite{Ishikawa:2007nn}, filled
 triangle is that of PACS-CS \cite{Kuramashi}, open symbols are
 perturbatively renormalized masses in Ref.~\cite{Ishikawa:2007nn}.
}
\label{fig:mud}
\end{center}
\end{figure}

As the last step we apply our renormalization factor to the bare PCAC
masses in Refs.~\cite{Ishikawa:2007nn,Aoki:2008sm,Kuramashi} to obtain
values for the renormalized quark masses.
%and see their scaling behavior.
The numerical results are given in Table \ref{tab:qmass}
and are plotted in Fig.~\ref{fig:mud} both for the averaged up and down
quark mass (left panel) and for the strange quark mass (right panel).
For the old CP-PACS/JLQCD work of Ref.~\cite{Ishikawa:2007nn}
results with $K$ and $\phi$ meson input are plotted (filled squares and
triangles) together with perturbatively renormalized masses using
tadpole improved renormalization factor (open squares and circles).
The triangle represents the result for the more recent work of PACS-CS
with the pion mass reaching down to $m_\pi=155$ MeV.

It is disappointing that the old CP-PACS/JLQCD results do not exhibit a
better scaling behavior by going from perturbative to non-perturbative
renormalization factor.
However, we should note a significant change in the average up and down
quark mass with the recent PACS-CS work (filled triangle).
This represents a systematic error due to chiral extrapolation of the
old CP-PACS/JLQCD work whose pion mass reached only $m_\pi\sim500$ MeV.
We should also note that the renormalization factor $Z_A$ involves a
large uncertainty at $\beta=1.83$ which is not reflected in the error
bar of Fig.~\ref{fig:mud}.
We feel that results at $\beta=2.05$ using simulation with physical pion
mass are needed to find the convincing values for light quark masses with
our approach.

%Scaling behavior seems not to be improved for previous work
%\cite{Ishikawa:2007nn} even if we adopt the non-perturbative
%renormalization factor.
%This may reflect a chiral symmetry breaking due to long chiral
%extrapolation rather than ${\cal O}(a^2)$ effect.

%The meson decay constants are renormalized non-perturbatively and
%results are shown in Figs.~\ref{fig:fpi} and \ref{fig:fv} for pseudo
%scalar and vector mesons.

\reseteqnum
%=============================================
\section{Conclusion}
%=============================================

We have presented a calculation of the quark mass renormalization factor
for the $N_f=2+1$ QCD in the mass independent Schr\"odinger functional
scheme in the chiral limit.
We adopt seven renormalization scales to cover from low to high energy
region.
Three lattice spacings are used to take the continuum limit.
%We adopt the same value for $(\beta,\kappa)$, which has been tuned
%for the previous work of the running coupling \cite{Aoki:2009tf}.
We calculate the pseudo scalar density renormalization factor and obtain
the SSF for the running mass.
Applying a perturbative improvement, we find that the step scaling
function shows a good scaling behavior and the continuum limit may be
taken safely with a constant extrapolation of the results for the finest
two lattice spacings.
The non-perturbative SSF turned out to be almost consistent with the
perturbative two loops result.

With the non-perturbative renormalization group flow we are able to
estimate the quark mass renormalization factor which renormalizes the
bare PCAC mass and give the RGI mass.
For this purpose we derive the renormalization factor of the pseudo
scalar density at low energy and that of axial vector current.
The results are used to renormalized the PCAC masses in the two recent
large scale simulations \cite{Ishikawa:2007nn,Aoki:2008sm,Kuramashi}.
We also evaluate the non-perturbative factor
$Z_m^{\ovl{\rm MS}}(\beta,\mu=2\ {\rm GeV})$ in $\ovl{\rm MS}$ scheme at
a scale $\mu=2$ GeV.

All statistical and systematic errors are included in the
renormalization factor $Z_M$ except for that due to the choice of the
reference scale $L_{\rm max}$.
A rough estimate of ${\cal O}((a/L_{\rm max})^2)$ effect suggests
possible 10 \%
%cannot deny a few percent 
systematic error at $\beta=1.90$.
However, a firmer conclusion requires the step scaling function at the
couplings stronger than those explored in the present work.
% we need to have the step scaling function at stronger coupling
%to get confident result, which should be a future work with better
%action to show a good scaling behavior.
We leave it as future work to examine this issue.

%The axial vector current renormalization factor seems to contain large
%lattice artifact at $\beta=1.83$.
%We may be better to perform lattice simulation at $\beta\ge1.90$ to take
%the continuum limit safely.
Applying our non-perturbative renormalization factor to the present
PACS-CS simulation with the pion mass as light as $m_\pi=155$ MeV yields
$m_{ud}^{\ovl{\rm MS}}(\mu=2\ {\rm GeV})=2.78(27)$ MeV for the average
up and down quark, and
$m_{s}^{\ovl{\rm MS}}(\mu=2\ {\rm GeV})=86.7(2.3)$ MeV for the strange
quark.
Simulations at weaker couplings under way will tell if these values stay
toward the continuum limit.

%
%=============================================
\section*{Acknowledgments}
%=============================================
%
This work is supported in part by Grants-in-Aid of the Ministry
of Education, Culture, Sports, Science and Technology-Japan
 (Nos. 18104005, 20105001, 20105002, 20105003, 20105005, 20340047,
 20540248, 21340049, 22105501, 22244018, 22740138).
Part of the calculations were performed by using the
RIKEN Integrated Cluster of Clusters facilities.

%=============================================

\clearpage
%%%%%%%%%%%%%%%%%%%%% table %%%%%%%%%%%%%%%%%%%%%%%%%%%%%%%%%%%%%%%%%%%%%%%%%%%%
\begin{table}[htb]
\begin{center}
\begin{tabular}{|c|c|c|c|c|c|c||c|}
\hline
$\beta$ & $\kappa$ & $L/a$ & $u=\ovl{g}^2(L)$ & $Z_P(g_0,L/a)$ &
$2L/a$ & $Z_P(g_0,2L/a)$ & $\Sigma_P(u,a/L)$ \\
\hline
$2.15747$ & $0.134249$ & $4$ & $3.4102(99)$ & $0.68631(56)$ &
$8$ & $0.5559(10)$ & $0.8095(17)$\\
$2.34652$ & $0.134439$ & $6$ & $3.415(16)$ & $0.64909(91)$ &
$12$ & $0.5464(22)$ & $0.8416(36)$\\
$2.5$ & $0.133896$ & $8$ & $3.418(19)$ & $0.64138(86)$ &
$16$ & $0.5469(19)$ & $0.8527(34)$\\
\hline
$2.5352$ & $0.132914$ & $4$ & $2.6299(29)$ & $0.75417(51)$ &
$8$  & $0.64852(95)$ & $0.8598(14)$\\
$2.73466$ & $0.133083$ & $6$ & $2.6292(77)$ & $0.71639(68)$ &
$12$ & $0.6349(14)$ & $0.8862(22)$\\
$2.9$ & $0.132658$ & $8$ & $2.6317(125)$ & $0.70575(63)$ &
$16$ & $0.6312(18)$ & $0.8944(28)$\\
\hline
$2.9605$  & $0.131831$ & $4$ & $2.1279(23)$ & $0.80163(44)$ &
$8$  & $0.71330(87)$ & $0.8898(12)$\\
$3.16842$ & $0.131997$ & $6$ & $2.1249(56)$ & $0.76386(50)$ &
$12$ & $0.69554(73)$ & $0.9104(12)$\\
$3.3$     & $0.131743$ & $8$ & $2.1289(92)$ & $0.74887(68)$ &
$16$ & $0.6808(14)$ & $0.9091(21)$\\
\hline
$3.33886$ & $0.131092$ & $4$ & $1.8426(19)$ & $0.82973(38)$ &
$8$  & $0.75240(85)$ & $0.9069(11)$\\
$3.55351$ & $0.131244$ & $6$ & $1.8375(32)$ & $0.79389(42)$ &
$12$ & $0.73496(68)$ & $0.9256(10)$\\
$3.7$     & $0.131021$ & $8$ & $1.8403(59)$ & $0.77923(55)$ &
$16$ & $0.7224(12)$ & $0.9271(17)$\\
\hline
$3.93653$ & $0.130195$ & $4$ & $1.5248(10)$ & $0.86074(35)$ &
$8$  & $0.79485(72)$ & $0.92348(92)$\\
$4.15042$ & $0.130356$ & $6$ & $1.5300(40)$ & $0.82826(38)$ &
$12$ & $0.77829(57)$ & $0.93955(81)$\\
$4.3$     & $0.1302$ & $8$ & $1.5242(35)$ & $0.81401(47)$ &
$16$ & $0.7652(11)$ & $0.9400(15)$\\
\hline
$4.74$   & $0.12934$ & $4$ & $1.24874(84)$ & $0.88808(28)$ &
$8$  & $0.83339(62)$ & $0.93842(76)$\\
$4.94755$& $0.129495$ & $6$ & $1.2483(15)$ & $0.85871(34)$ &
$12$ & $0.81476(41)$ & $0.94880(61)$\\
$5.1$    & $0.129376$ & $8$ & $1.2488(28)$ & $0.84501(40)$ &
$16$ & $0.80536(82)$ & $0.9531(11)$\\
\hline
$5.87312$& $0.128517$& $4$ & $0.99982(61)$ & $0.91197(27)$ &
$8$  & $0.86779(48)$ & $0.95152(60)$\\
$6.06879$& $0.12867$ & $6$ & $1.00130(97)$ & $0.88669(29)$ &
$12$ & $0.85142(41)$ & $0.96026(56)$\\
$6.2$    & $0.1286$  & $8$ & $1.0006(16)$  & $0.87496(24)$ &
$16$ & $0.84102(70)$ & $0.96121(84)$\\
\hline
\end{tabular}
\caption{The pseudo scalar density renormalization factor $Z_P$ at scale
 $L$ and $2L$.
 The value of $\beta$ and $\kappa$ has been tuned to reproduce the same
 physical box size $L$ and near zero PCAC mass in
 Ref.~\cite{Aoki:2009tf}.
 The step scaling function $\Sigma_P$ is also listed.}
\label{tab:beta-kappa}
\end{center}
\end{table}

\begin{table}[htb]
\begin{center}
\begin{tabular}{|c|c|c|c|c|c|c|c|}
\hline
$\beta$ & $\kappa$ & $L/a$ & \# of confs. & $m_{\rm PCAC}$
 & $2L/a$ & \# of confs. & $m_{\rm PCAC}$\\
\hline
$2.15747$ & $0.134249$ & $4$ & $19600$ & $0.03103(24)$&
$8$  & $12600$ & $0.03876(13)$\\
$2.34652$ & $0.134439$ & $6$ & $10300$ & $0.00264(18)$&
$12$ & $7600$ & $0.002856(78)$\\
$2.5$     & $0.133896$ & $8$ & $15200$ & $0.000753(79)$&
$16$ & $9800$ & $0.000517(35)$\\
\hline
$2.5352$  & $0.132914$ & $4$ & $20000$ & $0.02587(19)$ &
$8$  & $8000$ & $0.02808(11)$ \\
$2.73466$ & $0.133083$ & $6$ & $12000$ & $0.00251(10)$ &
$12$ & $8200$ & $0.001792(55)$\\
$2.9$     & $0.132658$ & $8$ & $24000$ & $0.000781(47)$&
$16$ & $8300$ & $0.000502(27)$\\
\hline
$2.9605$  & $0.131831$ & $4$ & $20000$ & $0.02179(15)$ &
$8$  & $8000$ & $0.021887(62)$\\
$3.16842$ & $0.131997$ & $6$ & $22000$ & $0.001917(74)$&
$12$ & $26800$ & $0.001093(20)$\\
$3.3$     & $0.131743$ & $8$ & $16000$ & $0.000479(49)$&
$16$ & $13200$ & $0.000280(17)$\\
\hline
$3.33886$ & $0.131092$ & $4$ & $20000$ & $0.01937(12)$ &
$8$  & $8000$ & $0.018782(68)$\\
$3.55351$ & $0.131244$ & $6$ & $20800$ & $0.002015(60)$&
$12$ & $24300$ & $0.001005(18)$\\
$3.7$     & $0.131021$ & $8$ & $16000$ & $0.000492(38)$&
$16$ & $12200$ & $0.000231(16)$\\
\hline
$3.93653$ & $0.130195$ & $4$ & $20000$ & $0.016138(93)$ &
$8$  & $8000$ & $0.015220(45)$\\
$4.15042$ & $0.130356$ & $6$ & $16000$ & $0.001450(47)$ &
$12$ & $23600$ & $0.000604(13)$\\
$4.3$     & $0.1302$   & $8$ & $16000$ & $-0.000088(31)$&
$16$ & $10200$ & $-0.000260(14)$\\
\hline
$4.74$   & $0.12934$  & $4$ & $20000$ & $0.013028(79)$&
$8$  & $8000$ & $0.011884(41)$\\
$4.94755$& $0.129495$ & $6$ & $16000$ & $0.001371(41)$&
$12$ & $35200$ & $0.0005863(92)$\\
$5.1$    & $0.129376$ & $8$ & $16000$ & $0.000243(25)$&
$16$ & $14100$ & $0.0000853(89)$\\
\hline
$5.87312$& $0.128517$& $4$ & $20000$ & $0.010385(62)$&
$8$  & $8000$ & $0.009065(33)$\\
$6.06879$& $0.12867$ & $6$ & $16000$ & $0.000910(31)$&
$12$ & $19200$ & $0.0002673(99)$\\
$6.2$    & $0.1286$  & $8$ & $32000$ & $0.000103(14)$&
$16$ & $11000$ & $-0.0000139(89)$\\
\hline
\end{tabular}
\caption{Number of configurations for each run.
Also listed is the PCAC mass at each parameter.}
% input of the normalization factor \eqn{eqn:normalization}.}
\label{tab:beta-kappa-conf}
\end{center}
\end{table}

\begin{table}[htb]
\begin{center}
\begin{tabular}{|c|c|c|c|c|c|c||c|}
\hline
$\beta$ & $\kappa$ & $L/a$ & $u=\ovl{g}^2(L)$ & $Z_P(g_0,L/a)$ &
$2L/a$ & $Z_P(g_0,2L/a)$ & $\Sigma_P(u,a/L)$ \\
\hline
$10$ & $0.1270893$ & $4$ & $0.58565(34)$ & $0.95007(21)$ &
$8$ & $0.92391(27)$ & $0.97247(36)$\\
$20$ & $0.1260654$ & $4$ & $0.29543(17)$ & $0.975596(93)$ &
$8$ & $0.96211(11)$ & $0.98618(15)$\\
$40$ & $0.1255571$ & $4$ & $0.14876(23)$ & $0.987683(80)$ &
$8$ & $0.981223(64)$ & $0.99346(10)$\\
$60$ & $0.1253871$ & $4$ & $0.099336(40)$ & $0.991832(46)$ &
$8$ & $0.9873514(42)$ & $0.995482(62)$\\
\hline
$10$ & $0.1272305$ & $6$ & $0.59707(44)$ & $0.93262(13)$ &
$12$ & $0.91141(31)$ & $0.97726(36)$\\
$20$ & $0.1261216$ & $6$ & $0.29775(33)$ & $0.96681(11)$ &
$12$ & $0.95557(13)$ & $0.98837(18)$\\
$40$ & $0.1255700$ & $6$ & $0.149219(45)$ & $0.983422(56)$ &
$12$ & $0.977882(57)$ & $0.994366(81)$\\
$60$ & $0.1253863$ & $6$ & $0.099664(47)$ & $0.988962(38)$ &
$12$ & $0.985302(42)$ & $0.996299(57)$\\
\hline
$10$ & $0.1272310$ & $8$ & $0.6051(12)$  & $0.92377(19)$ &
$16$ & $0.90192(77)$ & $0.97635(85)$\\
$20$ & $0.1261176$ & $8$ & $0.29971(49)$ & $0.962064(82)$ &
$16$ & $0.95088(29)$ & $0.98837(31)$\\
$40$ & $0.1255626$ & $8$ & $0.14948(27)$ & $0.981092(50)$ &
$16$ & $0.97552(10)$ & $0.99432(12)$\\
\hline
\end{tabular}
 \caption{The pseudo scalar density renormalization factor $Z_P$ at scale
 $L$ and $2L$ at very weak coupling region $\beta\ge10$.
 The step scaling function $\Sigma_P$ is also listed.}
\label{tab:pt-beta-kappa}
\end{center}
\end{table}

\begin{table}[htb]
\begin{center}
\begin{tabular}{|c|c|c|c|c|c|c|c|}
\hline
$\beta$ & $\kappa$ & $L/a$ & \# of confs. & $m_{\rm PCAC}$
 & $2L/a$ & \# of confs. & $m_{\rm PCAC}$\\
\hline
$10$ & $0.1270893$ & $4$& $9700$ & $0.005287(49)$ & 
$8$& $7200$ & $0.004447(20)$\\
$20$ & $0.1260654$ & $4$& $10000$ & $0.002057(23)$ & 
$8$& $7200$ & $0.001461(8)$\\
$40$ & $0.1255571$ &$4$&  $10000$ & $0.000401(14)$ &
$8$&  $7200$ & $-0.000084(4)$\\
$60$ & $0.1253871$ &$4$&  $10000$ & $-0.000128(8)$ &
$8$&  $7200$ & $-0.000558(3)$\\
\hline
$10$ & $0.1272305$ & $6$ & $20000$ & $0.000296(17)$&
$12$ & $8000$ & $-0.0001431(94)$\\
$20$ & $0.1261216$ & $6$ & $10000$ & $-0.000122(12)$&
$12$ & $7200$ & $-0.0004012(50)$\\
$40$ & $0.1255700$ & $6$ & $10000$ & $-0.000353(6)$&
$12$ & $7200$ & $-0.0005258(26)$\\
$60$ & $0.1253863$ & $6$ & $10000$ & $-0.000417(4)$&
$12$ & $7200$ & $-0.0005604(15)$\\
\hline
$10$ & $0.1272310$ & $8$ & $12000$ & $-0.000069(14)$&
$16$ & $3600$ & $-0.0001635(93)$ \\
$20$ & $0.1261176$ & $8$ & $12000$ & $-0.000216(7)$&
$16$ & $2700$ & $-0.0002723(67)$ \\
$40$ & $0.1255626$ & $8$ & $14400$ & $-0.000256(3)$&
$16$ & $2700$ & $-0.0002934(24)$ \\
\hline
\end{tabular}
 \caption{Number of configurations for each run at very weak coupling.
 Also listed is the PCAC mass at each parameter.}
\label{tab:pt-beta-kappa-conf}
\end{center}
\end{table}

\begin{table}[htb]
\begin{center}
\begin{tabular}{|c|c|c|c|c|}
\hline
$u$ & $\sigma_P(u)$ & $\Sigma_P(u,1/8)$ & $\Sigma_P^{\rm (2)}(u,1/6)$
 & $\Sigma_P^{\rm (2)}(u,1/4)$\cr
\hline
$1.0006$ & $0.96165(47)$ & $0.96121(84)$& $0.96185(56)$ & $0.96206(60)$ \cr
$1.2488$ & $0.95117(53)$ & $0.9531(11)$ & $0.95055(61)$ & $0.95177(77)$ \cr
$1.5242$ & $0.94113(74)$ & $0.9400(15)$ & $0.94136(82)$ & $0.94002(94)$ \cr
$1.8403$ & $0.92730(86)$ & $0.9271(17)$ & $0.9274(10)$  & $0.9272(11)$  \cr
$2.1289$ & $0.9113(10)$  & $0.9091(21)$ & $0.9119(12)$  & $0.9136(12)$  \cr
$2.6317$ & $0.8898(17)$  & $0.8944(28)$ & $0.8872(22)$  & $0.8898(14)$  \cr
$3.4178$ & $0.8474(25)$  & $0.8527(34)$ & $0.8411(36)$  & $0.8493(18)$  \cr
\hline
\end{tabular}
\caption{The perturbatively  improved SSF 
 $\Sigma_P^{\rm (2)}(u,a/L)$ at "two loop" level.
 The improvement is not applied for $a/L=1/8$.
 The SSF $\sigma_P(u)$ in the continuum is also listed, which is given by
 a constant fit of two data at finest lattice spacings $1/6$, $1/8$.}
\label{tab:ssf}
\end{center}
\end{table}

\begin{table}[htb]
\begin{center}
\begin{tabular}{|c|c|c|c|}
\hline
$\beta$ & $K$-input\cite{Ishikawa:2007nn}
 & $\phi$-input\cite{Ishikawa:2007nn} 
% & $K$-input\cite{Aoki:2008sm}
 & $K$-input\cite{Kuramashi}\\
\hline
 $1.83$ & $0.1174(23)$ & $0.1184(26)$ & \\
 $1.90$ & $0.0970(26)$ & $0.0971(25)$
% & $0.0907(13)$
 & $0.08995(40)$\\
 $2.05$ & $0.0701(29)$ & $0.0702(28)$ & \\
\hline
\end{tabular}
\caption{Lattice spacing $a$ in unit of fm from large scale simulations
 \cite{Ishikawa:2007nn} and 
\cite{Aoki:2008sm,Kuramashi}.}
\label{tab:lattice-spacing}
\end{center}
\end{table}

\begin{table}[htb]
\begin{center}
\begin{tabular}{|c|c|c|c|c|}
\hline
$\beta$ & $\kappa$ & $L_{\rm max}/a$ & $\ovl{g}^2(L_{\rm max})$
 & $Z_P(g_0,a/L_{\rm max})$ \\
\hline
$1.83$ & $0.13608455$ & $4$ & $5.565(54)$ & $0.57519(32)$ \\
\hline
$1.90$ & $0.1355968$ & $4$ & $4.695(23)$ & $0.60784(27)$ \\
\hline
$2.05$ & $0.1359925$ & $6$ & $4.740(79)$ & $0.56641(44)$ \\
\hline
\end{tabular}
 \caption{The renormalized coupling and the renormalization factor $Z_P$
 at $\beta=1.83$, $1.90$, $2.05$ for low energy scale $L_{\rm max}$.}
\label{tab:lmax}
\end{center}
\end{table}

\begin{table}[htb]
\begin{center}
\begin{tabular}{|c|c|c|c|c|c|c|c|c|c|c|}
\hline
 & size & $\beta$ & $\kappa$ & $\theta$ & traj. & $m_{\rm PCAC}$
 & $Z_V$ & $Z_A$ & $Z_A^{\rm con}$\\
\hline
&$6^3\times18$ & $1.83$ & $0.138466$ & $0.5$ & $149000$ & $-0.00065(67)$
% & $0.7538(17)$ & $0.9574(88)$ & $0.540(28)$ & $0.9590(88)$ & $0.556(30)$
 & $0.7533(16)$
% & $0.9605(87)$ & $0.537(22)$
 & $0.553(25)$ & $0.9622(88)$\\
&$6^3\times18$ & $1.83$ & $0.138466$ & $0$ & $22300$ & $0.0083(28)$
% & $0.7392(18)$ & $0.935(23)$ & $0.81(11)$ & $0.941(19)$ & $0.663(77)$
 & $0.7390(16)$
% & $0.932(21)$
 & $0.84(33)$ & $0.938(18)$\\
\hline
&$8^3\times18$ & $1.83$ & $0.138616$ & $0.5$ & $128000$ & $-0.00036(52)$
% & $0.71373(36)$ & $0.927(11)$ & $0.653(10)$ & $0.9314(90)$ & $0.655(10)$
 & $0.71343(32)$
% & $0.927(10)$ & $0.647(10)$
 & $0.650(10)$ & $0.9317(90)$\\
&$8^3\times18$ & $1.83$ & $0.138616$ & $0$ & $19200$ & $0.00081(97)$
% & $0.71507(84)$ & $0.953(16)$ & $0.641(18)$ & $0.948(14)$ & $0.629(16)$
 & $0.71522(71)$
% & $0.952(13)$ & $0.639(18)$
 & $0.626(15)$ & $0.947(13)$\\
\hline
&$10^3\times24$ & $1.83$ & $0.138640$ & $0.5$ & $5400$ & $0.0037(17)$
% & $0.7025(10)$ & $1.017(71)$ & $0.581(42)$ & $0.923(55)$ & $0.604(50)$
 & $0.7022(11)$
% & $0.589(38)$ & $1.025(65)$
 & $0.609(45)$ & $0.927(49)$\\
&$10^3\times24$ & $1.83$ & $0.138640$ & $0$ & $8800$ & $0.0009(10)$
% & $0.70769(81)$ & $0.982(34)$ & $0.592(24)$ & $0.966(25)$ & $0.577(22)$
 & $0.70768(66)$
% & $0.979(27)$ & $0.607(26)$
 & $0.594(24)$ & $0.963(23)$\\
\hline
$*$ & $12^3\times30$ & $1.83$ & $0.138700$ & $0.5$ & $3700$ & $-0.0006(23)$
% & $0.7022(11)$ & $0.97(12)$ & $0.590(65)$ & $1.01(14)$ & $0.604(71)$
 & $0.7013(14)$
% & $0.604(61)$ & $0.91(16)$
 & $0.619(70)$ & $0.95(17)$\\
\hline
\end{tabular}
 \caption{The (axial) vector current renormalization factor at
 $\beta=1.83$ for various box sizes.
 Two choices of $\theta=0.5$ and $\theta=0$ are adopted for comparison.
 We also listed $Z_A$ without disconnected diagrams as $Z_A^{\rm con}$.
 Data with star symbol is adopted for our $Z_A$.}
\label{tab:za-b183}
\end{center}
\end{table}

\begin{table}[htb]
\begin{center}
\begin{tabular}{|c|c|c|c|c|c|c|c|c|c|c|}
\hline
&size & $\beta$ & $\kappa$ & $\theta$ & traj. & $m_{\rm PCAC}$
 & $Z_V$ & $Z_A$ & $Z_A^{\rm con}$\\
\hline
&$8^3\times18$ & $1.90$ & $0.137556$ & $0.5$ & $88700$ & $0.00129(36)$
% & $0.74195(67)$ & $0.8568(57)$ & $0.733(17)$ & $0.8522(50)$ & $0.730(16)$
 & $0.74157(64)$
% & $0.8573(53)$ & $0.732(15)$
 & $0.729(15)$ & $0.8526(48)$\\
&$8^3\times18$ & $1.90$ & $0.137600$ & $0$ & $128000$ & $-0.00042(28)$
% & $0.74379(48)$ & $0.8521(96)$ & $0.7503(67)$ & $0.8526(92)$ & $0.7587(68)$
 & $0.74365(44)$
% & $0.8524(89)$ & $0.7495(76)$
 & $0.7578(69)$ & $0.8529(88)$\\
\hline
&$10^3\times24$ & $1.90$ & $0.137556$ & $0.5$ & $11300$ & $0.00256(66)$
% & $0.73303(68)$ & $0.893(17)$ & $0.714(22)$ & $0.875(15)$ & $0.725(22)$
 & $0.73321(49)$
% & $0.703(23)$ & $0.891(17)$
 & $0.716(24)$ & $0.873(15)$\\
$*$&$10^3\times24$ & $1.90$ & $0.137556$ & $0$ & $40800$ & $0.00191(41)$
% & $0.7358(38)$ & $0.892(16)$ & $0.825(23)$ & $0.876(14)$ & $0.785(20)$
 & $0.7354(37)$
% & $0.816(22)$ & $0.891(14)$
 & $0.781(20)$ & $0.874(13)$\\
\hline
&$12^3\times30$ & $1.90$ & $0.137503$ & $0.5$ & $4000$ & $0.00664(81)$
% & $0.73030(78)$ & $1.01(10)$ & $0.649(61)$ & $0.859(85)$ & $0.747(73)$
 & $0.73048(66)$
% & $0.633(60)$ & $1.00(10)$
 & $0.728(70)$ & $0.850(85)$\\
\hline
\end{tabular}
 \caption{The (axial) vector current renormalization factor at
 $\beta=1.90$ for various box sizes.
 Two choices of $\theta=0.5$ and $\theta=0$ are adopted for comparison.
 We also listed $Z_A$ without disconnected diagrams as $Z_A^{\rm con}$.
 Data with star symbol is adopted for our $Z_A$.}
\label{tab:za-b190}
\end{center}
\end{table}

\begin{table}[htb]
\begin{center}
\begin{tabular}{|c|c|c|c|c|c|c|c|c|c|}
\hline
&size & $\beta$ & $\kappa$ & traj. & $m_{\rm PCAC}$
 & $Z_V$ & $Z_A$ & $Z_A^{\rm con}$\\
\hline
&$8^3\times18$ & $2.05$ & $0.136116$ & $14400$ & $0.00305(40)$
% & $0.7840(10)$ & $0.8225(24)$ & $0.8217(36)$ & $0.8211(22)$ & $0.8144(43)$
 & $0.78369(99)$
% & $0.8226(20)$ & $0.8196(43)$
 & $0.8145(52)$ & $0.8211(22)$\\
\hline
$*$&$12^3\times30$ & $2.05$ & $0.136116$ & $5400$ & $0.00201(29)$
% & $0.77321(92)$ & $0.8157(27)$ & $0.7857(64)$ & $0.8122(28)$ & $0.808(11)$
 & $0.77314(82)$
% & $0.8157(25)$ & $0.7687(77)$
 & $0.804(15)$ & $0.8119(28)$\\
\hline
\hline
$*$&$8^3\times18$ & $3.0$ & $0.132420$ & $7200$ & $0.00070(14)$
% & $0.87132(94)$ & $0.8840(10)$ & $0.88862(67)$ & $0.88346(98)$ & $0.88565(69)$
 & $0.87109(91)$
% & $0.88385(90)$ & $0.88873(88)$
 & $0.88578(57)$ & $0.88335(95)$\\
\hline
$*$&$8^3\times18$ & $4.0$ & $0.130581$ & $8100$ & $-0.000032(58)$
%& $0.90506(82)$ & $0.91425(77)$ & $0.91418(84)$ & $0.91418(75)$ & $0.91397(85)$
 & $0.90495(82)$
% & $0.91427(74)$ & $0.91426(89)$
 & $0.91405(83)$ & $0.91420(74)$\\
\hline
$*$&$8^3\times18$ & $5.0$ & $0.129471$ & $5500$ & $-0.000060(66)$
% & $0.9246(13)$ & $0.93151(29)$ & $0.93183(40)$ & $0.93158(29)$ & $0.93179(42)$
 & $0.9245(13)$
% & $0.93149(29)$ & $0.93174(46)$
 & $0.93183(39)$ & $0.93154(31)$\\
\hline
\end{tabular}
 \caption{The (axial) vector current renormalization factor at
 $\beta=2.05$ and larger for $\theta=0.5$.
 We also listed $Z_A$ without disconnected diagrams as $Z_A^{\rm con}$.
 Data with star symbol is adopted for our $Z_A$.}
\label{tab:za-b205}
\end{center}
\end{table}

\begin{table}[htb]
\begin{center}
\begin{tabular}{|c|c|c|c|c|c|}
\hline
$\beta$
 & $Z_M$ & $Z_m^{\ovl{\rm MS}}$ ($K$)
 & PT(tad) $Z_m^{\ovl{\rm MS}}$ ($K$)
 & $Z_m^{\ovl{\rm MS}}$ ($\phi$)
 & PT(tad) $Z_m^{\ovl{\rm MS}}$ ($\phi$)\cr
\hline
$1.83$ \cite{Ishikawa:2007nn} & $1.33(15)$ & $1.04(12)$
 & $1.07161$
 & $1.04(12)$
 & $1.07019$\cr
$1.90$ \cite{Ishikawa:2007nn} & $1.693(46)$ & $1.315(35)$ 
 & $1.09973$
 & $1.315(35)$
 & $1.09955$\cr
$2.05$ \cite{Ishikawa:2007nn} & $1.862(41)$ & $1.417(29)$ 
 & $1.14487$
 & $1.416(29)$
 & $1.1446$\cr
\hline
$1.90$ \cite{Kuramashi} & $1.693(46)$ & $1.347(36)$
 & $1.11322$ & &\cr
\hline
\end{tabular}
\caption{$Z_M$ for the RGI mass and $Z_m^{\ovl{\rm MS}}(2\ {\rm GeV})$
 in the $\ovl{\rm MS}$ scheme, which are designed to renormalize the
 bare PCAC mass in two recent simulations of Ref.~\cite{Ishikawa:2007nn}
 and \cite{Aoki:2008sm,Kuramashi}.
 ($K$) or ($\phi$) means which meson mass is used for physical scale
 input.
 Perturbative results with tadpole improvement are also listed for $Z_m^{\ovl{\rm MS}}(2\ {\rm
 GeV})$.
 A slight difference in $Z_m^{\ovl{\rm MS}}$ at $\beta=1.90$ between
 Ref.~\cite{Ishikawa:2007nn} and \cite{Kuramashi} is due to
 that of the lattice spacing, from which physical unit is introduced.
}
\label{tab:zm}
\end{center}
\end{table}

\begin{table}[htb]
\begin{center}
\begin{tabular}{|c|c|c||c|c||c|c|}
\hline
$\beta$ & $\kappa$ & $L_{\rm max}/a$ & $\ovl{g}^2$ & $m_{\rm PCAC}$ &
$Z_P$ & $m_{\rm PCAC}$ \cr
\hline
$1.83$ & $0.138192$ & $6$ & $9.37(27)$ & $0.00231(59)$
 & $0.44601(86)$ & $0.00439(35)$ \cr
\hline
$1.90$ & $0.137289$ & $6$ & $6.60(12)$ & $0.00035(37)$
 & $0.49605(66)$ & $0.00313(22)$ \cr
\hline
$2.05$ & $0.13611599$ & $8$ & $6.01(21)$ & $0.00026(15)$
 & $0.51155(57)$ & $0.000733(74)$ \cr
\hline
\end{tabular}
\caption{The renormalized coupling $\ovl{g}^2$ and the renormalization
 factor $Z_P$ at larger box size of $L_{\rm max}/a$ at three $\beta$'s.
 Two PCAC masses are evaluated with different Dirichlet boundary
 condition.}
\label{tab:Lmax}
\end{center}
\end{table}

\begin{table}[htb]
\begin{center}
\begin{tabular}{|c|c|c|c|c|c|c|}
\hline
$\beta$ & $Z_M$ & $Z_m^{\ovl{\rm MS}}$ ($K$)\cr
\hline
$1.83$ \cite{Ishikawa:2007nn} & $1.46(18)$ & $1.15(14)$ \cr
$1.90$ \cite{Ishikawa:2007nn} & $1.837(57)$ & $1.418(41)$\cr
$2.05$ \cite{Ishikawa:2007nn} & $1.890(47)$ & $1.437(31)$\cr
\hline
$1.90$ \cite{Kuramashi} & $1.837(57)$ & $1.452(42)$ \cr
\hline
\end{tabular}
\caption{$Z_M$ and $Z_m^{\ovl{\rm MS}}(2\ {\rm GeV})$ evaluated with
 larger box size $L_{\rm max}$.}
\label{tab:zmLmax}
\end{center}
\end{table}

\begin{table}[htb]
\begin{center}
\begin{tabular}{|c|c|c|c|c|c|c|c|c|}
\hline
$\beta$
 & $M_{ud}^{\rm RGI}$ ($K$) & $m_{ud}^{\ovl{\rm MS}}$ ($K$)
 & $M_{s}^{\rm RGI}$ ($K$) & $m_s^{\ovl{\rm MS}}$ ($K$)
 & $M_{ud}^{\rm RGI}$ ($\phi$) & $m_{ud}^{\ovl{\rm MS}}$ ($\phi$)
 & $M_{s}^{\rm RGI}$ ($\phi$) & $m_s^{\ovl{\rm MS}}$ ($\phi$)\cr
\hline
$1.83$ \cite{Ishikawa:2007nn}
 & $3.30(38)$
 & $2.59(30)$
 & $86(10)$
 & $67.7(7.9)$
 & $3.31(39)$
 & $2.58(30)$
 & $99(11)$
 & $77.7(9.0)$
\cr
$1.90$ \cite{Ishikawa:2007nn}
 & $4.47(17)$
 & $3.47(13)$
 & $115.3(4.4)$
 & $89.5(3.4)$
 & $4.46(16)$
 & $3.46(12)$
 & $124.8(5.4)$
 & $96.9(4.2)$
\cr
$2.05$ \cite{Ishikawa:2007nn}
 & $5.29(24)$
 & $4.02(18)$
 & $136.1(6.5)$
 & $103.6(4.9)$
 & $5.29(24)$
 & $4.02(18)$
 & $143.0(8.1)$
 & $108.8(6.1)$
\cr
\hline
$1.90$ \cite{Kuramashi}
 & $3.49(34)$
 & $2.78(27)$
 & $109.0(3.0)$
 & $86.7(2.3)$
 & & & &\cr
\hline
\end{tabular}
\caption{Non-perturbatively renormalized mass for the averaged up and
 down quark and for the strange quark
 in the two recent simulations of
 Ref.~\cite{Ishikawa:2007nn,Aoki:2008sm,Kuramashi}.
 $M^{\rm RGI}$ is the RGI mass and $m^{\ovl{\rm MS}}$ is that in the
 $\ovl{\rm MS}$ scheme at a scale $\mu=2$ GeV.
 The unit is in MeV.
 ($K$) or ($\phi$) means which meson mass is used for physical scale
 input.
}
\label{tab:qmass}
\end{center}
\end{table}

\clearpage
%%%%%%%%%%%%%%%%%%% Figures %%%%%%%%%%%%%%%%%%%%

%\clearpage
%\begin{figure}[htbp]
%\begin{center}
% \includegraphics[width=7.cm]{fig/fpi.eps}
% \includegraphics[width=7.cm]{fig/fk.eps}
% \caption{Scaling behavior of non-perturbatively renormalized $f_\pi$
% (left) and $f_K$ (right); filled circles are from CP-PACS/JLQCD result
% \cite{Ishikawa:2007nn}, open circles are the same data with
% perturbative renormalization factor.
% Filled triangle is that of PACS-CS \cite{Kuramashi}.
% }
%\label{fig:fpi}
%\end{center}
%\end{figure}
%
%\begin{figure}[htbp]
%\begin{center}
% \includegraphics[width=7.cm]{fig/fv.eps}
% \caption{Non-perturbatively renormalized $f_\rho$ (circle), $f_{K^*}$
% (square) and $f_\phi$ (diamond).
% All data are from PACS-CS result \cite{Kuramashi}.
% }
%\label{fig:fv}
%\end{center}
%\end{figure}

\end{document}